\documentclass[times]{article}

\usepackage[usenames, dvipsnames]{color} 
\usepackage{amsmath}          
\usepackage[utf8x]{inputenc}  
\usepackage{graphicx}         
\usepackage{amssymb,amsfonts}
\usepackage{fixmath}
\usepackage{multirow}
\usepackage{pgfplots}
\usepackage{bbold}
\usepackage[ruled,vlined]{algorithm2e}
\usepackage{algorithmicx,algpseudocode}
\usepackage{tikz}
\usepackage[labelsep=quad,indention=10pt]{caption}
\usepackage{bm}
\usepackage{tabu}
\usepackage{float}
\usepackage{subfig}

\usepackage{amssymb}

\def\TT{\textsf{T}}

\newcommand\bG{\mathbf{G}}
\newcommand\bPhi{\mathbf{\Phi}}
\newcommand\bLambda{\mathbf{\Lambda}}

\newcommand\bx{\bm{h}}
\newcommand\bh{\bm{h}}

\newcommand\tegam{{T_{\gamma}}}
\newcommand\tegamk{{T_{\gamma}^{k}}}
\newcommand{\nMC} {\ensuremath{\texttt{n}_{\texttt{MC}}}}
\newcommand{\SRS} {\ensuremath{\texttt{}_{\texttt{SRS}}}}
\newcommand{\OK} {\ensuremath{\texttt{}_{\texttt{OK}}}}
\newcommand{\PRS} {\ensuremath{\texttt{}_{\texttt{PRS}}}}
\newcommand{\nd} {\ensuremath{\texttt{n}_{\texttt{d}}}}
\newcommand{\ns} {\ensuremath{\texttt{n}_{\texttt{s}}}}
\newcommand{\nf} {\ensuremath{\texttt{n}_{\texttt{f}}}}
\newcommand{\nii} {\ensuremath{\texttt{n}_{i}}}
\newcommand{\ngg} {\ensuremath{\texttt{n}_{\gamma}}}
\newcommand{\nG} {\ensuremath{\texttt{n}_{\texttt{G}}}}
\newcommand\RR{\leavevmode\hbox{$\rm I\!R$}}

\newcommand{\PDMText}[1]{{\color{black} #1}}
\newcommand{\AGGText}[1]{{\color{black} #1}}
\newcommand{\MRAText}[1]{{\color{black} #1}}

\begin{document}
\title{Nonintrusive Uncertainty Quantification for automotive crash problems with VPS/Pamcrash}

\author{Marc Rocas(1,2), Alberto Garc\'ia-Gonz\'alez(1), Sergio Zlotnik(1,3), \\Xabier Larr\'ayoz(2), Pedro D\'iez(1,3)\\ \\
$1$- Laboratori de C\`alcul Num\`eric, E.T.S. de Ingenier\'ia de Caminos,\\ Universitat Polit\`ecnica de Catalunya -- BarcelonaTech\\
$2$- SEAT, Martorell, Barcelona\\
$3$- The International Centre for Numerical\\ Methods in Engineering, CIMNE, Barcelona}

\maketitle 

\begin{abstract}
Uncertainty Quantification (UQ) is a key discipline for computational modeling of complex systems, enhancing reliability of engineering simulations. In crashworthiness, having an accurate assessment of the behavior of the model uncertainty allows reducing the number of prototypes and associated costs. Carrying out UQ in this framework is especially challenging because it requires highly expensive simulations. In this context, surrogate models (metamodels) allow drastically reducing the computational cost of Monte Carlo process. Different techniques to describe the metamodel are considered, Ordinary Kriging, Polynomial Response Surfaces and a novel strategy (based on Proper Generalized Decomposition) denoted by Separated Response Surface (SRS). A large number of uncertain input parameters may jeopardize the efficiency of the metamodels. Thus, previous to define a metamodel, kernel Principal Component Analysis (kPCA) is found to be effective to simplify the model outcome description. A benchmark crash test is used to show the efficiency of combining metamodels with kPCA.\\

Keywords: Uncertainty Quantification, 
Crashworthiness,
Separated Response Surface (SRS), 
kernel Principal Component Analysis (kPCA), 
Kriging, 
Surrogate modeling.

\end{abstract}

\section{Introduction}\label{sec:intro}
Uncertainty Quantification (UQ) in crashworthiness simulations is becoming an important asset to verify the design of vehicle structures. An important challenge for automotive industry is guaranteeing safety and reducing costs, and this requires virtually testing a large number of tentative designs, and assess the dispersion of the results of the virtual model due to the uncertain input data. UQ is particularly relevant for the crashworthiness analysis, where many uncertainties have to be tackled together and then propagated in the simulation of this extremely complex, nearly chaotic, phenomenon. 

The growth and universal accessibility to computational resources and the robustness of codes, offer a good perspective on the possibilities of producing UQ for complex problems. However, in the case of crashworthiness simulations, a single simulation takes up to 18 CPU hours in a High Performance Computing facility. Thus, the very large number of queries associated with a standard UQ process are practically unaffordable in this context. One viable alternative is using a surrogate model (or metamodel) build upon a reduced number of full-order simulations (denoted as \emph{training set}), see \cite{qiu2018crashworthiness,wang2018crashworthiness,moustapha2014metamodeling} for different approaches and comparative analyses. Still, the viability of metamodels is limited by the number of input parameters: a large number of parameters results in a highly multidimensional input space and therefore the engineer is afflicted by the so-called \emph{curse of dimensionality}. \AGGText{If the parametric model is already low-dimensional (3-4 design parameters), the actual threat is not the \emph{curse of dimensionality} but dimensionality reduction is still necessary to computationally afford the simulation process for decision making . This is  common in crashworthiness and in the example included here for illustration.} 

The idea is to determine a low number of relevant parameters (as combinations of the original ones) properly representing all the variability of the dataset. Principal Component Analysis (PCA) is the standard dimensionality reduction technique, to be used if the data structure is such that the low-dimensional subset where the data is contained (also referred as manifold) is linear. Other \emph{manifold learning} techniques identify nonlinear low-dimensional structures. Among them, kernel Principal Component Analysis (kPCA) is considered here because it is one the simplest approaches, see  \cite{garcia2020kernel} for a synthetic presentation. The combination of dimensionality reduction with surrogate modeling is a common strategy to carry out UQ in different disciplines and contexts \cite{lataniotis2018extending,li2020efficient,nagel2017uncertainty}.

This paper analyzes the combination of different alternatives for dimensionality reduction techniques and surrogate models for UQ in crashworthiness simulations with uncertain input parameters. Among the different techniques explores, the novel combination of kPCA and Separated Response Surface (SRS) demonstrates interesting properties. Other strategies are also used to define surrogate model fitting the training set like Polynomial Regression Surface (PRS) and Ordinary Kriging (OK). \AGGText{To do this, we used Monte Carlo sampling (as it is the simplest method for statistical analyses) in two steps of the proposed methodology: 1) to obtain the training set for dimensionality reduction and surrogate model reconstrucction, and 2) once the low dimensional surrogate models were properly developed, standard Monte Carlo is performed (practically with no computational cost) to increase the probabilistic resolution in the description of the QoI. }

This paper is structured as follows: section \ref{sec:Benchmark} presents the benchmark problem for crashworthiness. The proposed UQ methodology is described in section \ref{sec:DRM}, which is divided in the three subsections. First, the main ideas of the kPCA technique are recalled. Next, the three different surrogates under consideration are briefly introduced (SRS, OK and PRS). Finally, the different surrogates are readily used to quantify the uncertainty of the output. 
Section \ref{sec:App_benchmark_crash_problem} illustrates how the methodologies presented in the previous section perform for the crash problem proposed in Section \ref{sec:Benchmark}. Finally, section \ref{sec:conclusions} includes some concluding remarks.

\section{Benchmark problem for crash}\label{sec:Benchmark}

\begin{figure}[h]
\begin{center}
\includegraphics[width=0.70\textwidth]{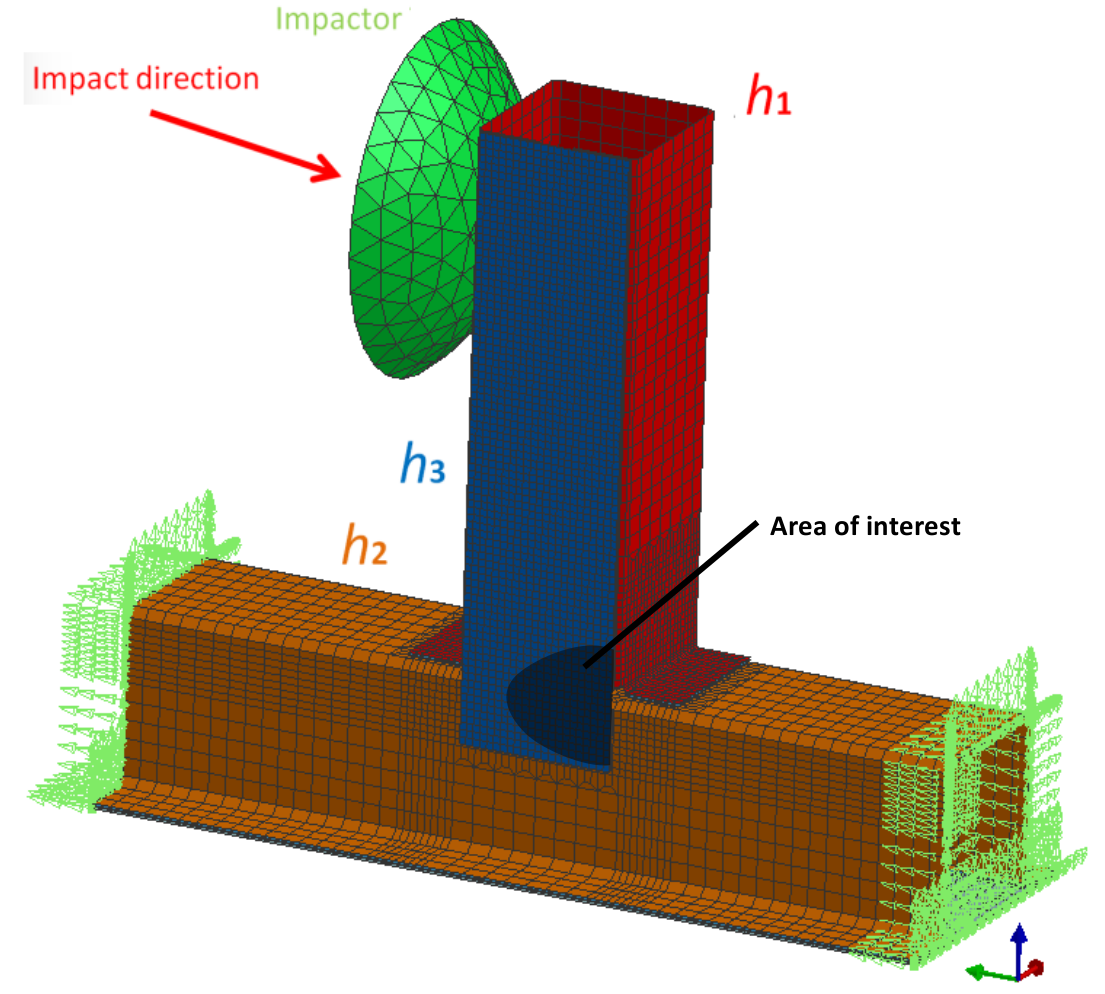}
\caption{Crash benchmark. Thicknesses $h_1$, $h_2$ and $h_3$ are the three input random parameters corresponding to the vertical profile (red), horizontal profile (orange) and plate profile(blue). The impactor (green), and the area of elements of interest (black) are also depicted.}
\label{fig:mesh}
\end{center}
\end{figure}

The benchmark crash problem under consideration is illustrated in Fig. \ref{fig:mesh}. It corresponds a reduced test model, ideally reproducing the main characteristics of the simulation of a B-pillar, a well-known structural component of cars. This particular benchmark test is used for different research studies in the Volkswagen Group. For the sake of saving computational cost and time, this model is often used to test new materials, adhesives, welding spots or other conditions because, due to the simplicity of the model, the numerical response requires a computation of approximately 20 minutes. \MRAText{Even so, computing the solutions of the final training set with 2366 samples used here required around 788 hours (approximately 32 days) of computational time in one of the SEAT clusters. Besides, the computational time required for a standard full crash model is around one day per simulation. Thus, any effort in devising strategies to build a reliable training set with the minimum number of full-order solutions is worthwhile}.

The driving force in the model is provided by the impactor (green zone in Fig. \ref{fig:mesh}), that crashes at a speed of 50 mm/s against the vertical profile (red zone) during one second. \MRAText{This velocity might seem to be low, nevertheless this is a standard benchmark test of interest for the Volkswagen Group accounting for rate effects in the model. The reason is that the high deformations are concentrated locally in the structure in specific areas and times, so that local rate effects are significant.}

The three structural parts are plates made of laminated steel sheet manufactured by cold folding. All the parts are joined with a structural adhesive bond, its material properties are characterized by Volkswagen. 

\MRAText{The structure is modelled using the Belytschko-Tsay shell element (which is a standard option in the VPS/Pamcrash package, very popular in the automotive sector, in particular for SEAT engineers) with one integration point in the plane.} The impactor is considered to be a rigid body. The complete model has a total of $13 908$ nodes (with 6 degrees of freedom).

The Quantity of Interest (QoI) to be analyzed is the final plastic strain in the 142 shell elements of the area depicted in black  in Fig. \ref{fig:mesh}.

The numerical solver is implemented in VPS/Pamcrash \cite{Pamcrash}, with the shell finite element discretization mentioned above and an explicit time stepping scheme to solve the dynamical problem. The displacements of the points at the ends of the horizontal profile are prescribed to zero (points marked with green arrows in  Fig. \ref{fig:mesh}). The contact between the different components of the structure are treated with the surface-surface model defined in VPS/Pamcrash.

Thicknesses $h_1$, $h_2$ and $h_3$ of the three parts of the structure are considered to be stochastic parameters, that is random variables collected in vector $\bh=[h_1,h_2,h_3]^{\TT}$. Their aleatory character is associated with the imperfections produced during the manufacturing process. Random variables $h_1$, $h_2$ and $h_3$ are assumed to be normal and uncorrelated, that is $h_i \thicksim \mathcal{N}(\mu_i,\sigma_i^2)$ and $\text{cov}(h_{i}, h_{j})=0$, for $i,j=1,2,3$. \MRAText{In each of the three parts, the corresponding thickness is considered to be constant. Besides, the three thicknesses $h_1$, $h_2$ and $h_3$ are modelled as having the same mean $\mu_1=\mu_2=\mu_3=1.2\,\text{mm}$ and standard deviation $\sigma_1=\sigma_2=\sigma_3=0.12\,\text{mm}$.}

In order to build a training set, and as a first assessment of the stochastic behaviour of the system, a number of $\ns=2366$ Monte Carlo realizations (or samples) are performed. Thus, $\ns$ values of the input parameters $\bh^{i}$, for $i=1,2,\dots,\ns$ are generated with a random number generator and the corresponding VPS/Pamcrash solutions are obtained. These solutions (in particular the vectors containing the plastic strain in the $d=142$ elements of the zone of interest) are collected in a training set matrix $\mathbf{X}=[\mathbf{x}^1 \mathbf{x}^2 \cdots \mathbf{x}^{\ns}]\in \RR^{d \times \ns}$, where each column $\mathbf{x}^{i}=[x_{1}^{i} \dots x_{d}^{i}]^{\TT}$ is the VPS/Pamcrash solution corresponding to input $\bh^{i}$.
The actual QoI is the average plastic strain in the zone, is represented by a form $l^{0}(\cdot)$, and for each $\bh^{i}$ and  $\mathbf{x}^{i}$ reads 
$$
l^{0}(\mathbf{x}^{i})=\frac{1}{d} \sum_{j=1}^{d}x_j^i .
$$
Note that the Monte Carlo process with $\ns=2366$ samples is considered here as a reference, and it is only obtained in the academic example under consideration. The number of full-scale computations affordable for a real problem in the automotive industrial practice is much lower. A methodology to select the optimal number of samples (the lowest providing some insights in the assessment of uncertainty) is discussed in section \ref{sec:App_benchmark_crash_problem}.

The points constituting the training are often taken as realizations of random variables (Monte Carlo) as indicated above. They can be also taken as pseudo random (e.g. Latin hypercube) or deterministic (e.g. Hammersley points, Halton sequences) \cite{wong1997sampling}.

\section{Dimensionality reduction, surrogate model and UQ}\label{sec:DRM}
The number of samples $\ns$ affordable in a real problem is generally not sufficient to produce a proper Monte Carlo assessment of the statistical properties of the output of the system. A review of nonintrusive UQ methodologies for crashworthiness, see \cite{rocas2020nonintrusive}, demonstrates that the standard Monte Carlo sampling is extremely demanding and, in practice, beyond the possibilities of standard industrial practitioners. 

\begin{figure}[h]
\begin{center}
\includegraphics[width=0.9\textwidth]{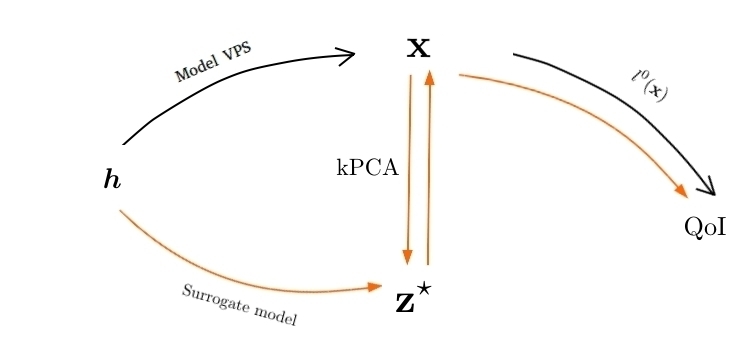}
\caption{Schematic illustration of the methodology.}\label{fig:methodology}
\end{center}
\end{figure}

As indicated in the previous section, the standard Monte Carlo approach consists in generating random samples of the input, running the model and retrieving statistics of the output (or any QoI). This is what corresponds to the upper part (black arrows) in the scheme of Fig.\ref{fig:methodology}. 

However, the part of the standard model (also denoted as full-order, here computed with VPS/Pamcrash) is too computationally expensive to be performed for the number of samples providing statistical relevance. Thus, the alternative is to replace this full-order model by a surrogate, that is a simple functional transformation from $\bh$ to $\mathbf{x}$. The surrogate is created using a training set consisting in data generated by the full-order model. 

An additional difficulty is encountered due to the high-dimension of the outcome of the model, $\mathbf{x}$. It is complicated to create a high-dimensional functional approximation having a target space of $d$ (here 142) dimensions. Thus, previous to undertake the determination of the surrogate, it is convenient to apply some dimensionality reduction technique. In the context of crashworthiness simulation, the data generated by the models are often adopting nonlinear data structures \cite{garcia2020kernel,van2009dimensionality}. Thus, it is expected to require nonlinear dimensionality reduction (kPCA). 

\PDMText{The QoI is introduced as an essential indicator for decision making. The QoI summarizes the information contained in $\mathbf{x}$. Quantifying the uncertainty of the QoI is sufficient to take some decisions. For instance, to verify the crashworthiness response of the structural design.  
Uncertainty Quantification of high-dimensional objects like $\mathbf{x}$ is cumbersome and the outcome is difficult to use as a tool supporting decision making. In that sense, the stochastic assessment focuses in a low-dimensional (even purely scalar) QoI, rather than in a high-dimensional object like   $\mathbf{x}$. However, a deeper analysis of the phenomenon requires understanding the underlying mechanisms associated with the overall mechanical response of the system. In that sense, all the information contained in $\mathbf{x}$ is pertinent. The fact that the model order reduction strategy is able to recover back the full-order object in as accurately as possible is therefore extremely advantageous. In this aspect, kPCA behaves much better than PCA in many cases: the simple QoI is fairly approximated by the PCA reduction, but kPCA improves the mapping back to the original  variable $\mathbf{x}$.}

All these aspects are covered in the methodology described in remainder of the paper, having the following steps:
\begin{itemize}
\item[$\bullet$]  \textbf{Creation of training set. }  Generate $\ns$ realizations of the input parameters $\bh^{i}$, for $i=1,2,\dots,\ns$, and compute the corresponding full-order solutions $\mathbf{x}^{i}$ (that constitute the training set). 
\item[$\bullet$] \textbf{Dimensionality reduction. } Analyze the training set and find the principal components allowing to reduce the dimensionality of the family of solutions. In practice, this boils down to apply PCA or kPCA and determine a mapping between the solutions $\mathbf{x}\in \RR^{d}$ and some new variable $\mathbf{z}^{\star}\in\RR^{k}$ in a much lower-dimensional space ($k\ll d$). The mapping between $\mathbf{x}$ and $\mathbf{z}^{\star}$ is to be characterized forward and backward. \PDMText{The kPCA backward mapping is found to be more accurate than with PCA. That is, kPCA recovers with much more accuracy a full-order $\mathbf{x}$ associated with a reduced-order coordinate $\mathbf{z}^{\star}$. Although this advantage is often not perceptible when assessing a low-dimensional (or scalar) QoI, a proper  $\mathbf{x}$ recovery is crucial to deepen in the mechanical interpretation of the results. For instance, to identify the mechanisms associated with the different modes of the probability distribution.}
\item[$\bullet$] \textbf{Surrogate model. } The functional dependence $\mathbf{z}^{\star} = F(\bh)$ is determined from the data provided by the training set, and the dimensionality reduction. 
\item[$\bullet$]  \textbf{Complete Monte Carlo UQ (using surrogate). } Once the surrogate $F(\cdot)$ is available, for each input value $\bh$, the corresponding \textbf{\MRAText{$\mathbf{z^{\star}}$}} is straightforwardly computed as $F(\bh)$. Then the backward mapping produces the corresponding $\mathbf{x}$, and $l^{0}(\mathbf{x})$ is the associated QoI. The concatenation of the three operations is computationally affordable. Therefore standard Monte Carlo can be performed with a sufficient number of realizations.
\end{itemize}

\MRAText{The different aspects of the devised methodology are described in detail in the following sections. It is important noting that, among the four conceptual steps mentioned above, the computational cost is concentrated in the creation of the training set. Obtaining this representative collection of solutions requires a computational time in the range of weeks or months, depending on the type of simulation in crashworthiness. The other steps: dimensionality reduction, surrogate modeling and Monte Carlo UQ represent in practice a negligible amount of computational efforts (in the order of seconds).}

\subsection{Dimensionality reduction}\label{sec:DR} 

The training set matrix $\mathbf{X}=[\mathbf{x}^1 \mathbf{x}^2 \dotsb \mathbf{x}^{\ns}]\in \RR^{d \times \ns}$ is seen as a set of $\ns$ points in a $d$-dimensional space. The idea of dimensionality reduction is to find a subspace of lower dimension $k\ll d$ where the set of points is contained.

The PCA strategy consists in \emph{diagonalizing} the square $d\times d$ matrix $\mathbf{X}^{\TT} \mathbf{X}$ (covariance matrix), that is finding $\mathbf{U} \in \RR^{d \times d}$ such that 
\begin{equation}\label{eq:DiagXTX}
\mathbf{X}^{\TT} \mathbf{X} = \mathbf{U} \bLambda  \mathbf{U}^{\TT}
\end{equation}
where $\bLambda$ is a diagonal matrix with eigenvalues $\lambda_{1} \ge \lambda_{2} \ge \dots \ge \lambda_{d}$.
The dimension is reduced from $d$ to $k$ if the last $d-k$ eigenvalues are negligible with respect to the $k$ first. In this case, the new variable selected is 
\begin{equation}\label{eq:forwardMapping}
\mathbf{z}^{\star} = \mathbf{U}^{\star \, \TT}  \mathbf{x},
\end{equation}
being $\mathbf{U}^{\star} \in \RR^{d \times k}$ the matrix with the first $k$ columns of $\mathbf{U}$. Eq. \eqref{eq:forwardMapping} describes the forward mapping, that is how to map the high-dimensional vector $\mathbf{x}$
into the element $\mathbf{z}^{\star}$ reduced dimensional space, from dimension $d$ to dimension $k$. The backward mapping goes in the opposite direction and reads 
\begin{equation}\label{eq:backwardMapping}
\mathbf{x} = \mathbf{U}^{\star}  \mathbf{z}^{\star} .
\end{equation}

Thus, PCA is a straightforward methodology relying in the fact that the training set is lying in a linear subspace. In many cases, the structure of the low-dimensional manifold where the solution ranges is nonlinear and more sophisticated dimensionality reduction techniques are required. The kPCA is an alternative reformulation based on the traditional PCA, in fact, it performs PCA in a new feature space where the data is transformed from the original space \cite{scholkopf1998nonlinear}. 

The main characteristics of the kPCA are explained in detail in \cite{garcia2020kernel} and summarized here.
It is assumed that some transformation $\bPhi$ from $\RR^{d}$ to a higher-dimensional space is able to \emph{flatten} the training set. That is the transformed training set 
$\left \{ \bPhi(\mathbf{x}^1),   \bPhi(\mathbf{x}^2), \dots,  \bPhi(\mathbf{x}^{\ns})\right\}$
is such that PCA is able to discover a linear subspace of dimension $k$. In other words, the transformation $\bPhi$ maps the nonlinear manifold (of dimension $k$) where the training set ranges into a linear subspace.

The transformation $\bPhi$ that produces this effect is a priori unknown. However, it is worthy trying with some different alternatives and see what is the reduced dimension they propose: the best choice for $\bPhi$ is the one producing the lower value of $k$. Moreover, in practice $\bPhi$ is indirectly characterized using the \emph{kernel trick}. Thus, instead of describing directly $\bPhi$, an expression for the bivariate form $\kappa(\cdot,\cdot)$ is provided, assuming that the following relation between $\kappa(\cdot,\cdot)$ and $\bPhi(\cdot)$ holds
\begin{equation}\label{eq:kernelDef}
\kappa(\mathbf{x}^{i},\mathbf{x}^{j})=  \bPhi(\mathbf{x}^{i})^{\TT} \bPhi(\mathbf{x}^{j})
\end{equation}
for $i,j=1,2,\dots,\ns$.

A classical choice for the kernel $\kappa(\cdot,\cdot)$ is the so-called Gaussian kernel, defined as
\begin{equation}\label{eq:kernelGauss}
\kappa(\mathbf{x}^{i},\mathbf{x}^{j})=  \exp(-\beta \left\Vert \mathbf{x}^{i} - \mathbf{x}^{j} \right\Vert^{2})
\end{equation}
where $\beta$ is a parameter that, in the applications of this paper, is taken equal to 0.1.

Having the kernel at hand, one may compute a matrix equivalent to $\mathbf{X} \mathbf{X}^{\TT}$ for the samples transformed by $\bPhi$. This matrix is denoted by $\bG\in \RR^{\ns \times \ns}$ and has generic coefficient
\begin{equation}\label{eq:Gmatrix}
\left[ \bG \right]_{ij}=\kappa(\mathbf{x}^{i},\mathbf{x}^{j})
\end{equation}

It is worth noting that the eigenvalues of $\mathbf{X} \mathbf{X}^{\TT}$ are the same of those of $\mathbf{X}^{\TT} \mathbf{X}$, which are the ones extracted in \eqref{eq:DiagXTX}. Actually, $\bG$ is also readily diagonalized and the following factorization is obtained
\begin{equation}\label{eq:DiagXXT}
\bG  = \mathbf{V} \tilde{\bLambda}  \mathbf{V}^{\TT}
\end{equation}
where $ \tilde{\bLambda}$ contains the same non-zero eigenvalues that would be obtained from diagonalizing the corresponding covariance matrix, which is not available.

Thus, the eigenvalues $\lambda_{1} \ge \lambda_{2} \ge \dots \ge \lambda_{\ns}$ are computed, and the reduced dimension $k$ is selected such that the last $\ns-k$ eigenvalues are negligible with respect to the $k$ first. 

Once $k$ is obtained, the original variable $\mathbf{x}\in \RR^{d}$ is mapped into a variable in the reduced space, 
$\mathbf{z}^{\star}\in \RR^{k}$ using the following the expression
\begin{equation}\label{eq:forwardMappingkPCA}
\mathbf{z}^{\star} = \mathbf{V}^{\star \, \TT}   \mathbf{g}(\mathbf{x}),
\end{equation}
where $\mathbf{V}^{\star} \in \RR^{\ns\times k}$ is the matrix with the first $k$ columns of $\mathbf{V}$ and 
$ \mathbf{g}(\mathbf{x})\RR^{\ns}$ is a vector with generic component
\begin{equation}\label{eq:gVector}
 \left[\mathbf{g}(\mathbf{x})\right]_{i}=\kappa(\mathbf{x}^{i},\mathbf{x})
\end{equation}
for $i=1,\dots,\ns$.

As described in detail in \cite{garcia2020kernel}, if the samples transformed by $\bPhi$ are not centred, some corrections have to be done and both matrix $\bG$ and vector $\mathbf{g}$ have to be modified accordingly. These corrections are straightforward and are omitted here for the sake of a simpler presentation.

Equations \eqref{eq:forwardMappingkPCA} and \eqref{eq:gVector} characterize the forward kPCA mapping, from $\mathbf{x}$ to $\mathbf{z}^{\star}$. The backward mapping for kPCA is not as simple as for the PCA version described in equation \eqref{eq:backwardMapping}. A point $\mathbf{z}^{\star}$ in the reduced space is mapped back to a point $\mathbf{x}$ which is recovered as a weighted average of the points of the training set, namely
\begin{equation}\label{eq:backwardMappingkPCA}
\mathbf{x} = \sum_{i=1}^{\ns} w_i(\mathbf{z}^{\star}) \mathbf{x}^{i}  
\,\,\, \text{ , with weights  such that } \sum_{i=1}^{\ns} w_i(\mathbf{z}^{\star})= 1
\end{equation}
The weights $w_i(\mathbf{z}^{\star})$ are computed such that the forward mapping of $\mathbf{x}$ is as close as possible to 
$\mathbf{z}^{\star}$. A popular strategy to compute these weights with a simple approach is to use a radial basis interpolation concept based on the distances of $\mathbf{z}^{\star}$ to the images of the sample points, $\mathbf{z}^{\star \, i}$ for $i=1,\dots,\ns$. That is computing $d_{i}=\Vert \mathbf{z}^{\star} - \mathbf{z}^{\star \, i}  \Vert$ and taking any value of $w_i(\mathbf{z}^{\star})$ decreasing with $d_{i}$, for example
$w_i(\mathbf{z}^{\star}) \propto \frac{1}{d_{i}^{2}}$. 
%
\subsection{Surrogate model}\label{sec:SM}
As already announced, the dimensionality reduction techniques presented above are a previous step to build a surrogate model. The training set is now used to approximate the functional dependence associated with the full-order model. The final goal is to 
compute $\mathbf{x}$ as an easy-to-evaluate function of $\bh$, that is the surrogate. With the dimensionality reduction, this is split in two steps: a surrogate from $\bh$ to $\mathbf{z}^{\star}$ plus the backward mapping from $\mathbf{z}^{\star}$ to $\mathbf{x}$, see Fig. \ref{fig:methodology}.

Here, the surrogates are presented as generic methodologies to establish a functional dependency among some input $\bh$ and some output function $\mathbf{y}(\bh)$ (we use $\mathbf{y}$ to account for any output, that could be either $\mathbf{x}$ or $\mathbf{z}^{\star}$). Obviously, doing the surrogate with $\mathbf{z}^{\star}$ has the advantage of dealing with a much lower dimension (number of components) of the model output. In practice, for the sake of a simpler presentation, a scalar output $Y(\bh)$ is considered in the following, that stand, for example, for any of the components of $\mathbf{y}(\bh)$. In the examples, $Y$ coincides with the first component of the reduced space using kPCA, that is $Y=[\mathbf{z}^{\star}]_{1}$.

%

In the following, the parameters describing the stochastic input space where the function takes values are collected in the vector $\bh=[h_{1} \dots  h_{\nd}]^{\TT} \in \RR^{\nd}$. Where $\nd$ is the number of stochastic dimensions of the problem ($\nd=3$ in the benchmark under consideration).

Thus, the goal is to approximate the functional dependence $Y(\bh)$ using the images of the points of the training set  $y^{k} = Y(\bh^{k})$, $k=1,\dots,\ns$, where . All the sample points are collected in the vector $\mathbf{y}=[y^1y^2...y^{\ns}]^T$. 


\subsubsection{Separated Response Surface (SRS)}

The idea of SRS is to find a separated approximation $F(\bh)$  to $Y(\bh)$. The separated character of $F(\bh)$ means that it is a sum of rank-one terms, being each rank-one term the product of sectional modes (the adjective sectional is used to indicate that the mode depends only on one of the parameters).  The algorithm employed to compute the SRS is based on the ideas of the least-squares PGD approximations described in detail in \cite{ComptesRendus20,diez2019encapsulated,lu2018multi,lu2018adaptive}.

Thus, $F(\bh)$ reads
\begin{equation}\label{eq:scattPGDform}
F(\bh)=\sum_{j=1}^{\nf} \sigma_{j} \prod_{i=1}^{\nd} f_{i}^{j} (h_{i})
\end{equation}
where each sectional mode $ f_{i}^{j} (h_{i})$ is represented in some discrete sectional space.
The discrete sectional space is generated by a family of functions \newline
$\left\{ \Psi^{i}_{1}(h_{i}) \, \Psi^{i}_{2}(h_{i})  \dots   \Psi^{i}_{\nii}(h_{i})  \right\}$
being $\nii$ the dimension of the sectional function space.
Accordingly, sectional modes have the following expression
\begin{equation}\label{eq:sectionalApprox}
f_{i}^{j} (h_{i}) = \sum_{m=1}^{\nii} a_{i}^{m} \Psi^{i}_{m}(h_{i})
\end{equation}
where the unknown coefficients $a_{i}^{m} $, for $i=1,\dots,\nd$ and $m=1,\dots,\nii$, have to be computed to determine the sectional mode $f_{i}^{j} (h_{i})$, for $j=1,2,\dots$

Different alternatives are available as the approximation space defined by $\Psi^{i}_{m}(h_{i})$. Here we have considered 
a finite element discretization with $\nii$ nodes (in 1D domains, $\nii-1$ elements).

A least-squares criterion based on a discrete Euclidean product is chosen to select $F(\bh)$. Thus, $F(\bh)\approx Y(\bh)$ is taken such that it minimizes 
\begin{equation}\label{eq:LSnorm}
\Vert F(\bh) - Y(\bh) \Vert^{2}=\left<  F-Y,F-Y\right>= \sum_{k=1}^{\ns} w^{k} (F(\bh^{k}) - y^{k})^{2} 
\end{equation}
where the weights $w^{k}$ are introduced to assimilate the sum into an integral, that is, to assume that 
$$
\int_{\Omega_{\bh}} F(\bh)\; d\bh \approx \sum_{k=1}^{\ns} w^{k} F(\bh^{k}) 
$$
note that the associated scalar product $\left< \cdot,\cdot \right>$ of two arbitrary functions $F$ and $G$ reads
\begin{equation}\label{eq:LSscalarProd}
\left<  F , G \right> = \sum_{k=1}^{\ns} w^{k} F(\bh^{k}) G(\bh^{k}). 
\end{equation}
\PDMText{Note that  weights $w^{k}$, $k=1,2,\dots,\ns$ must be selected corresponding to a quadrature having as integration points $\bh^{k}$, where $Y$ is known.
Typically, the distribution of points $\bh^{k}$ is provided by a stochastic sampling and cannot be enforced a priori by the user to obtain his/her preferred quadrature (e.g. a Gauss-Legendre quadrature or a composite Simpson's rule). Thus, the weights of the quadrature are adapted to optimize the integration order in a (multidimensional) Newton-Cotes fashion.  
}

Thus, least-squares solution in a linear functional space $\bm{V}$ is readily computed as a projection, that is finding $F\in\bm{V}$ such that
\PDMText{
\begin{equation}\label{eq:GlobalProj}
\left<  F , F^{\star} \right> = \left<  Y , F^{\star} \right> \text{ for all } F^{\star} \in \bm{V}
\end{equation}
Note that \emph{integral} equation \eqref{eq:GlobalProj} has to be fulfilled for any weighting function (or test function) $F^{\star}$ in $\bm{V}$, as in the standard weak form of a boundary value problem.}

The key aspect of any PGD algorithm is how to solve the rank-one approximation. That, is how to find an approximation to $Y(\bx)$ with a function of the form 
\begin{equation}\label{eq:scattPGDformRO}
F(\bx)=\sigma \prod_{i=1}^{\nd} f_{i} (h_{i})
\end{equation}
which is a particular case of \eqref{eq:scattPGDform} with just one term.

The standard PGD strategy consists in an alternate direction approach, that is to compute the sectional mode $f_{\gamma}$, the rest of the sectional modes $f_{i}$ for $i\neq \gamma$  are assumed to be known. Thus, in practice, $F(\bx)$ and $F^{\star} (\bx)$ are taken as
\begin{equation}\label{eq:RoFanddF1}
F(\bx)=f_{\gamma} (x_{\gamma})  \left[ \prod_{i\neq \gamma} f_{i} (h_{i}) \right]
\end{equation}
and
\begin{equation}\label{eq:RoFanddF2}
F^{\star} (\bx)= f^{\star}_{\gamma}(h_{\gamma}) \left[ \prod_{i\neq \gamma} f_{i} (h_{i})\right].
\end{equation}
\PDMText{This alternate directions strategy leads to a sectional problem, reduced to the $\gamma$ coordinate. The family of sectional problem is to be solved sequentially for  $\gamma=1,2,\dots,\ns$, and then iterated  until convergence is reached.}

For the sake of simplifying the writing, the computable term depending on all the sectional modes but $\gamma$ is denoted as $\tegam$ and $\tegamk$ when evaluated in $\bx^{k}$, namely
\begin{equation} \label{eq:tegam}
\tegam:=\prod_{i\neq \gamma} f_{i} (h_{i})
\text{ and }
\tegamk:=\prod_{i\neq \gamma} f_{i} (h_{i}^{k}).
\end{equation}

Thus, the sectional counterpart of \eqref{eq:GlobalProj} reads
\begin{equation}\label{eq:SectionalProj1}
\left< f_{\gamma} \tegam
, f^{\star} _{\gamma} \tegam
\right> = \left<  Y , f^{\star} _{\gamma}  \tegam \right> \text{ for all }f^{\star} _{\gamma} 
\end{equation}
that is
\begin{equation}\label{eq:SectionalProj2}
 \sum_{k=1}^{\ns} w^{k}  (\tegamk)^{2} f_{\gamma}(x_{\gamma}^{k}) 
f^{\star} _{\gamma} (h_{\gamma}^{k}) 
 =  \sum_{k=1}^{\ns} w^{k}  \tegamk y^{k}  f^{\star} _{\gamma} (x_{\gamma}^{k}).
 \end{equation}

Using a particular case of \eqref{eq:sectionalApprox}, that is 
\begin{equation}\label{eq:sectionalApprox2}
f_{\gamma} (h_{\gamma}) = \sum_{m=1}^{\ngg} a_{\gamma}^{m} \Psi^{\gamma}_{m}(h_{i})
 \end{equation}
 in \eqref{eq:SectionalProj2} and taking 
$ f^{\star} _{\gamma} (h_{\gamma}^{k}) =  \Psi^{\gamma}_{\ell}(h_{\gamma}^{k})$ for all $\ell=1,\dots,\ngg$ yields
\begin{equation}\label{eq:SectionalProj3}
\sum_{k=1}^{\ns} w^{k}  (\tegamk)^{2} \sum_{m=1}^{\ngg} a_{\gamma}^{m} \Psi^{\gamma}_{m}(h_{\gamma}^{k}) 
 \Psi^{\gamma}_{\ell}(h_{\gamma}^{k})
 =  \sum_{k=1}^{\ns} w^{k}  \tegamk y^{k}   
 \Psi^{\gamma}_{\ell}(h_{\gamma}^{k})
\end{equation}
or 
\begin{equation}\label{eq:SectionalProj3}
\sum_{m=1}^{\ngg} 
\underbrace{\left[ \sum_{k=1}^{\ns} w^{k}  (\tegamk)^{2}   \Psi^{\gamma}_{m}(h_{\gamma}^{k}) 
 \Psi^{\gamma}_{\ell}(h_{\gamma}^{k}) \right]}_{M_{\ell \, m}}
 a_{\gamma}^{m}
 =  \underbrace{\sum_{k=1}^{\ns} w^{k}  \tegamk y^{k}   
 \Psi^{\gamma}_{\ell}(h_{\gamma}^{k})}_{f_{\ell}}
\end{equation}
for all  $\ell=1,\dots,\ngg$. 
That is, a linear system of $\ngg$ equation with $\ngg$ unknowns 
\begin{equation}\label{eq:SectionalLinSystem}
\bm{M} \bm{a}_{\gamma} = \bm{f}.
\end{equation}

Once the sectional approximation is obtained solving \eqref{eq:SectionalLinSystem}, the loop in alternate directions iterations is continued until convergence and completion of the rank-one computation. As usual in PGD \cite{garikapati2020proper}, once the rank-one solution is obtained, the greedy approach aims at computing the next term (next $j$ in \eqref{eq:scattPGDform}).

\PDMText{As it is standard in this type of strategies, in order to compute an approximation having the separated form given in \eqref{eq:scattPGDform}, there are three nested loops. First, the greedy approach (loop in $j$) aims at computing rank-one terms having the form given in \eqref{eq:scattPGDformRO}. Then an alternated direction iterative scheme is applied consisting in two nested loops: the iterative loop to reach convergence (not described explicitly in this text with an iteration index) and an inner loop for $\gamma=1,2,\dots \ns$, ranging all sectional dimensions. This is standard in the references describing any PGD scheme, see \cite{diez2019encapsulated} for an algorithmic description.}

\PDMText{As mentioned above, functions $\Psi^{i}_{m}$ in \eqref{eq:sectionalApprox} are chosen as classical ${\cal{C}}^{0}$ finite elements shape functions. Contrary to other choices (e.g. high-order polynomials) these type of functions are more stable due to their local support but introduce a lack of smoothness (jumps in the first derivatives, singularities in the second derivatives). Consequently, when}
using a finite element approximation for the sectional modes, it is important having the possibility of enforcing the smoothness of the solution. This is equivalent to penalize in system  \eqref{eq:SectionalLinSystem}, the non-smoothness of the sectional function described in \eqref{eq:sectionalApprox2}. 

This requires, for instance, penalizing some postprocessed quantity of the sectional mode $f_{\gamma} (h_{\gamma})$, represented by the vector of nodal values, $\bm{a}_{\gamma}$. The quantity to be penalized, the lack of smoothness,  is represented by a matrix $\bm{G}$ mapping the nodal values of $f_{\gamma} (h_{\gamma})$ into the postprocessed quantity in some representative points. In the following,  $\bm{G}$ is taken as the standard \emph{gradient} operator, computing the derivatives of $f_{\gamma} (h_{\gamma})$  in the integration points of the elements of the mesh. Thus,  $\bm{G}$ is a $\nG \times \ngg$ matrix, being $\nG$ the number of integration points in the mesh (assuming that the dimension of the sectional space is 1). Thus, the measure of the lack of smoothness that has to be reduced is given by $ \bm{a}_{\gamma}^{\TT} \bm{G}^{\TT} \bm{G} \bm{a}_{\gamma}$. Provided that system  \eqref{eq:SectionalLinSystem} is equivalent to minimize the following functional
$$
\frac{1}{2} \bm{a}_{\gamma}^{\TT} \bm{M} \bm{a}_{\gamma} - \bm{f}^{\TT} \bm{a}_{\gamma} 
$$
Enforcing the smoothness requires minimizing the perturbed functional
$$
\frac{1}{2} \bm{a}_{\gamma}^{\TT} \bm{M} \bm{a}_{\gamma} - \bm{f}^{\TT} \bm{a}_{\gamma} + \lambda \frac{1}{2} \bm{a}_{\gamma}^{\TT} \bm{G}^{\TT} \bm{G} \bm{a}_{\gamma}
$$
for some value of the factor $\lambda$ that states the importance of the smoothing. The larger is $\lambda$, the smoother is the recovered solution. This results in the following linear system, which is a modification of \eqref{eq:SectionalLinSystem} accounting for the smoothing
\begin{equation}\label{eq:SectionalLinSystem2}
\left[ \bm{M} + \lambda \bm{G}^{\TT} \bm{G}  \right] \bm{a}_{\gamma} = \bm{f}
\end{equation}

\subsubsection{Ordinary Kriging (OK)}\label{subsec:OK}
\begin{figure}[h!]
  \centering
  \includegraphics[width=0.70\textwidth]{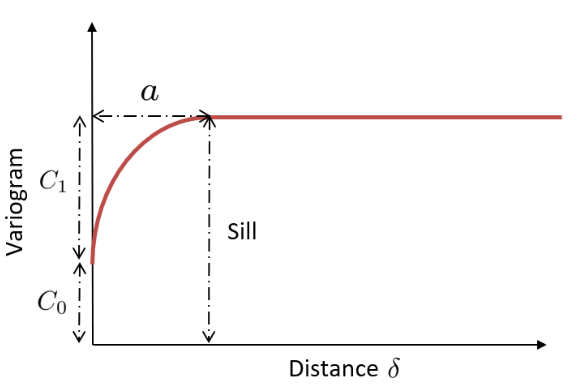}
  \caption{Variogram with the three main parameters. The nugget $C_0$, the range $a$ and the sill $C_0+C_1$.}
  \label{variogram}
 \end{figure}

Ordinary Kriging (OK) is an interpolation technique commonly used in engineering and originated for geostatistical problems \cite{oliver2014tutorial}. The OK method determines weights for a set of simulation points to calculate a prediction of a new sample. The weights are calculated with a variogram model that has the main feature to estimate variances for any distance. The kriging metamodel $F(\bh)$ of any point $\bh$ is defined by:
\begin{equation}
F (\bh)= \sum_{i=1}^{\ns} w_i y^i,
\end{equation}
where the unknowns $w_{i}$ are the weights and $y^i$ are the scalar values of the function to be interpolated. To determine the optimal values for the kriging weights, the variogram function plays an important role. There exist different variograms: Gaussian, exponential, linear among others \cite{oliver2014tutorial}. The OK matrix system to obtain the weights reads,
\begin{equation*}
 \left(
  \begin{array}{ccccc} 
   \gamma_{11} & \gamma_{12} & \ldots & \gamma_{1\ns} & 1\\
   \vdots & \vdots & \ddots & \vdots & \vdots \\
   \gamma_{\ns1} & \gamma_{\ns2} & \ddots & \gamma_{\ns \ns} & 1\\
   1&1&\ldots&1&0\\
  \end{array}
  \right)
  \left(
  \begin{array}{c} 
   w_{1} \\
   \vdots \\
   w_{\ns} \\
   \mu\\
  \end{array}
  \right)
    =
    \left(
  \begin{array}{c} 
   \gamma_{10} \\
   \vdots \\
   \gamma_{\ns0} \\
   1\\
  \end{array}
  \right).
\end{equation*}
 \PDMText{A specific condition for OK with respect to other Kriging methods is enforcing the sum of weights equal to 1, $\sum_{i=1}^{\ns}w_i=1$. This condition is achieved by introducing the new unknown $\mu$ as Lagrange multiplier  \cite{malvic2009linearity}.}
The entries of the matrix in the equation above depend on the variogram function $\gamma$ evaluated for each distance $\delta$ between a pair of samples, that is $\gamma_{ij}=\gamma(\delta)$, being $\delta=\Vert \bh^{i} - \bh^{j} \Vert$. The entries $\gamma_{10} \ldots \gamma_{\ns0}$ are evaluations of the variogram $\gamma$ between all the sample points with respect to the new (current) point. Here, we used the spherical variogram defined as:
\begin{equation}
\gamma(\delta) = \left\lbrace
\begin{array}{ll}
\textup{$C_0+C_1\left[ \frac{3}{2}\left(\frac{\delta}{a}\right)-\frac{1}{2}\left(\frac{\delta}{a}^3\right) \right],$}& 0 < \delta\leq a\\
\textup{$C_0+C_1,$}& \delta > a
\end{array}
\right.
\end{equation}
$C_0$ is the nugget constant representing the noise of the data, $a$ is the range of the transition zone  where the variogram levels off and the sill $(C_0+C_1)$ is defined as the total variance of the model. For the benchmark problem $C_0=0$, in consequence $C_1$ is the total variance of the model. In Fig.\ref{variogram} it is illustrated a spherical variogram function. 

\subsubsection{Polynomial Response Surface}\label{subsec:PR}
Polynomial Response Surface (PRS) has been applied in numerous studies to build  metamodel for different engineering problems \cite{gano2006comparison,fang2005comparative,giunta1998comparison}. 
It consists in a simple multidimensional polynomial fitting.
A second order polynomial model $F(\bh)$ takes the form,
\begin{equation}\label{eq:PRS}
F(\mathbf{h})= c_0+\sum_{i=1}^{\nd}c_i h_i+\sum_{i=1}^{\nd}c_{ii} h_i^2+\sum_{i=1}^{\nd-1} \sum_{j=i+1}^{\nd} c_{ij}h_ih_j,
\end{equation}
where $h_i$ is the $i$-th stochastic input, the different coefficients $c$ are the unknowns to be computed, collected in a vector  $\mathbf{c}$.  If the approximation was able to interpolate the data, the following linear system should be solved:
\begin{equation}\label{eq:AcZ}
\mathbf{A}\mathbf{c}=\mathbf{y},
\end{equation}
where $\mathbf{A}$ is the matrix containing the values of the different interpolation functions in \eqref{eq:PRS}, that is, for $\nd=3$,
$$
\left\{ 1, h_{1}, h_{2}, h_{3},  (h_{1})^{2}, (h_{2})^{2}, (h_{3})^{2}, h_{1}h_{2}, h_{1} h_{3}, h_{2} h_{3} \right\}
$$
in the sample points $\bh^{k}$, for $k=1,\dots,\ns$. This results in 
\begin{equation}
 \mathbf{A}= 
  \left[
  \begin{array}{cccccccccc} 
      1 & h^{1}_{1}  &h^{1}_{2} & h^{1}_{3} & \left(h^{1}_{1}\right)^{2} &\left(h^{1}_{2}\right)^{2} 
    & \left(h^{1}_{3}\right)^{2} & h^{1}_{1} h^{1}_{2}  & h^{1}_{1} h^{1}_{3} &h^{1}_{2} h^{1}_{3}\\
        1 & h^{2}_{1}  &h^{2}_{2} & h^{2}_{3} & \left(h^{2}_{1}\right)^{2} &\left(h^{2}_{2}\right)^{2} 
    & \left(h^{2}_{3}\right)^{2} & h^{2}_{1} h^{2}_{2}  & h^{2}_{1} h^{2}_{3} &h^{2}_{2} h^{2}_{3}\\
  \vdots &\vdots &\vdots&\vdots&\vdots&\vdots&\vdots&\vdots&\vdots&\vdots\\
    1 & h^{\ns}_{1}  &h^{\ns}_{2} & h^{\ns}_{3} & \left(h^{\ns}_{1}\right)^{2} &\left(h^{\ns}_{2}\right)^{2} 
    & \left(h^{\ns}_{3}\right)^{2} & h^{\ns}_{1} h^{\ns}_{2}  & h^{\ns}_{1} h^{\ns}_{3} &h^{\ns}_{2} h^{\ns}_{3}
  \end{array}
  \right],
\end{equation}
The model is often non-interpolative, with more equations (points in the sample) than unknowns (number of coefficients in $\mathbf{c}$). Therefore system \eqref{eq:AcZ} cannot be solved exactly but using a least squares minimization criterion, that is minimizing the Euclidean norm of the residual, namely $\Vert \mathbf{y}-{A}\mathbf{c}\Vert$. This results in taking the vector of unknown coefficients solution of
\begin{equation}
(\mathbf{A}^T\mathbf{A}) \mathbf{c}= \mathbf{A}^T\mathbf{y}.
\end{equation}
This method presents drawbacks for high-dimensional data and data with oscillations. Increasing the order of the Polynomials may  improve accuracy. However, for high-order approximations Runge's phenomenon creates instabilities and wrong predictions \cite{boyd2009divergence}.

\subsection{Uncertainty quantification}\label{sec:UncertaintyQuantification}

\subsubsection{Monte Carlo sampling with surrogate modeling}
Once the surrogate model is available, the Monte Carlo UQ assessment with a large number of samples $\nMC$  is produced at an affordable computational cost. 

Thus, for each of the three metamodels introduced above (SRS, OK and PRS) $\nMC$ realizations $\bh^{1},\bh^{2},\dots,\bh^{\nMC}$ are produced and the corresponding value of of the mean, variance (and standard deviation) and Probability Density Function (PDF, to be approximated as a histogram) is readily estimated:
\begin{equation}
\text{Mean}= \mathbb{E}[F(\bx)]= \frac{1}{\nMC}\sum_{k=1}^{\nMC}F(\bh^k)
\end{equation}
\begin{equation}
\text{Variance}= \sigma^{2}=\frac{1}{\nMC-1}\sum_{k=1}^{\nMC}\left(F(\bh^k)-\mathbb{E}[F(\bx)]\right)^2,
\end{equation}
The PDF corresponding to $Y=F(\bh)$ is denoted by $f_Y(y)$ and it is approximated by histogram $p_Y(y)$ computed on the basis of the $\nMC$ Monte Carlo samples $y^k=F(\bh^k)$, for $k=1,2,\dots,\nMC$. Note that histogram $p_Y(y)$ is a piecewise constant function defined over a partition in uniform intervals of the $Y$ domain, $\Omega_Y=\bigcup_{\ell = 1}^{n_Y} I_\ell $. Piecewise constant function $p_Y$  is such that for $y\in I_\ell$, $p_Y(y)$ is equal to the number of samples $y^k$ lying in $I_\ell$ divided by $\ns$.

Each response surface is used to generate the images of the $\nMC=50000$ samples of the input space $\bh \in \RR^{\nd}$, that is $\bh \rightarrow Y$). The backward mapping technique ($Y \rightarrow \mathbf{X} \rightarrow QoI$) described in Section \ref{sec:DR} is used to obtain the statistics of the QoI.

Comparing the obtained values of mean and variance is straightforward because they are scalar values. However, comparing PDFs is not as trivial. Here, the Kullback Leibler divergence technique is  proposed as a criterion to compare the PDF functions.

\subsubsection{Comparative criterion for PDFs}
Kullback-Leibler (KL) divergence is used as a comparative criterion for the different resulting Monte Carlo PDFs of each metamodel. This quantity measures differences between two PDF functions \cite{galas2017expansion}.  \PDMText{Two random variables $F$ and $G$ have PDFs $f$ and $g$. The KL divergence is introduced as a distance that quantifies if the two random variables are similar enough. The random variable $F$ and its PDF  $f$ are taken as the reference and $g$ is considered to be an approximation to $f$.}
Thus, KL divergence between the two continuous PDFs $f$ and $g$ reads:
\begin{equation}\label{eq:KLDivCont}
D_{KL}(F \Vert G)=\int_{-\infty}^{\infty}f(y) \, \log\left(\frac{f(y)}{g(y)}\right) \, dy.
\end{equation}
\PDMText{Note that equation \eqref{eq:KLDivCont} is associated with the notion of \emph{entropy} and it is interpreted as the relative entropy or the \emph{information gain} from $G$ to $F$.}

 \PDMText{In the case the PDFs are replaced by their discrete counterparts, that is histograms, instead of $f$ and $g$, one has histograms $p_{Y}$ and $q_{Y}$ with the format described in the previous section. Thus,  $p_{Y}$ and $q_{Y}$ are expressed as the values of the probability of being in each of the $n_{Y}$ bins, that is $p_{Y}(y^{\ell})$ and $q_{Y}(y^{\ell})$, for $\ell=1,2,\dots,n_{Y}$ and $y^\ell \in I_{\ell}$. The discrete counterpart of equation \eqref{eq:KLDivCont} reads}
\begin{equation}\label{eq:KLDivDisc}
D_{KL}(p_Y \Vert q_Y)=\sum_{\ell = 1}^{n_Y} p_Y(y^\ell) \, \log\left(\frac{p_Y(y^\ell)}{q_Y(y^\ell)}\right).
\end{equation}

 \PDMText{The values obtained using the discrete KL divergence introduced in equation \eqref{eq:KLDivDisc} depend on the number of bins, $n_{Y}$. In order to normalize these values, a normalizing constant is introduced providing a reference to understand wether the resulting discrete KL divergence is actually small enough. Note that the KL divergence is seen as a distance but it is not conceived as the norm of a difference. Thus, it is not possible to normalize dividing directly by the norm of $p_{Y}$ (or $f$ in \eqref{eq:KLDivCont}). In order to obtain a reference value, we propose taking the \emph{distance} of $p_{Y}$ to the less informative distribution, that is the uniform histogram $q_{U}$ such that  $q_{U}(y^{\ell})=\frac{1}{n_{Y}}$, for $\ell=1,2,\dots,n_{Y}$. 
The rationale behind this choice is taking $q_{U}$ as the \emph{zero} or absolute reference distribution. 
This value is denoted as $D_{KL}^{0}$ and reads
\begin{equation}\label{eq:KLDivDisc0}
D_{KL}^{0}=D_{KL}(p_Y \Vert q_U)=\sum_{\ell = 1}^{n_Y} p_Y(y^\ell) \, \log \left( n_{Y} \, p_Y(y^\ell) \right).
\end{equation}
Note that the  quantity $D_{KL}^{0}$ defined in  equation \eqref{eq:KLDivDisc0}  is actually the entropy of $p_{Y}$. Dividing the figures obtained with the KL divergence of \eqref{eq:KLDivDisc} by $D_{KL}^{0}$ provides a relative value that allows elaborating a more informed criterion to decide if $p_{Y}$ and $q_{Y}$ are sufficiently close to each other.
}

\section{Application to the crash problem}\label{sec:App_benchmark_crash_problem}

In this section, the combination methodology of kPCA + SRS is applied for the benchmark crash problem. Also, OK and PRS are applied to compare the SRS results.

\subsection{Sampling size for kPCA}\label{subsec:sampling_size}

\begin{figure}[h!]
  \centering
  \includegraphics[width=1\textwidth]{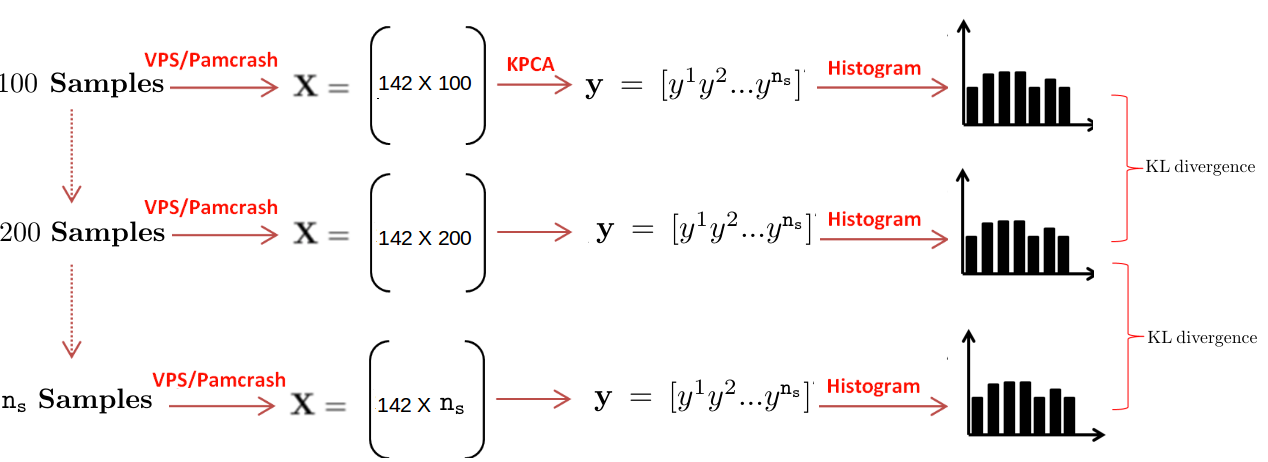}
  \caption{Flowchart scheme to select the number of simulation $\ns$ for the kPCA input matrix $\mathbf{X}$.}\label{ns_approach}
 \end{figure}

\AGGText{Initially, a certain amount of VPS/Pamcrash FE simulations is required to reconstruct the input matrix $\mathbf{X}$ for kPCA manifold analysis and dimensionality reduction (the training set computed in an offline phase).
These simulations are the initial samples for the data analysis. A key issue is quantifying the number of samples required to obtain enough and credible information to describe the low-dimensional manifold containing the solution. 
As described above, we devise the combined use of kPCA to reduce the dimensionality, different techniques to build a response surface, and the KL divergence as a measure to compare the different probability distributions and stop enriching the sampling.
This process is illustrated in the flowchart scheme shown in Fig.\ref{ns_approach}, and it is detailed next:}

\begin{figure}[h!]
  \centering
  \includegraphics[width=0.70\textwidth]{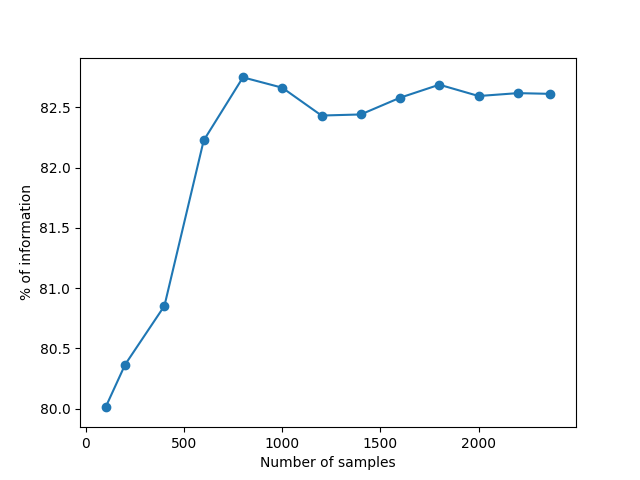}
  \caption{Quantity of information $[\%]$ stored by the first eigenvector by increasing the number of samples $\ns$.}\label{ns_info}
 \end{figure}

\AGGText{The eigenvalues of the kernel matrix $\bG  = \mathbf{V} \tilde{\bLambda}  \mathbf{V}^{\TT}$ in kPCA (equation \ref{eq:DiagXXT}), show the quantity of information collected by each associated eigenvector, as explained in section \ref{sec:DR} and more detailed in \cite{garcia2020kernel}. Summarizing, the largest eigenvalue measures the largest amount of information collected by the corresponding eigenvector. For instance, for the first eigenvalue of matrix $\tilde{\bLambda}$, the associated eigenvector is the first column of matrix $\mathbf{V}$. Adopting the first component as reduced model (keeping only one principal component, the first one), the $d = 142$ dimensions of the training set samples (each column of matrix $\mathbf{X}$) are reduced to one scalar number  and the $\ns$ samples are stored in the vector $\mathbf{y}=[y^1y^2...y^{\ns}]^T$, see sections \ref{sec:DR} and \ref{sec:SM} for more details.

In the case of our benchmark crash problem, reducing to one dimension collects more than $80\%$ of information of the manifold where data belong. This 80\% figure allows considering as admissible in this context, and in agreement with the resulting approximations, the very advantageous reduction to a single dimension. This figure is calculated along the sampling refinement process (for different values of $\ns$) to check the behaviour of the quantity of information retained in the one-dimensional reduction. This is shown in Fig.\ref{ns_info}, where it can be noticed that even using only 100 samples, the first eigenvalue collects already $80\%$ of information.}

\begin{figure}[h!]
  \centering
  \includegraphics[width=0.9\textwidth]{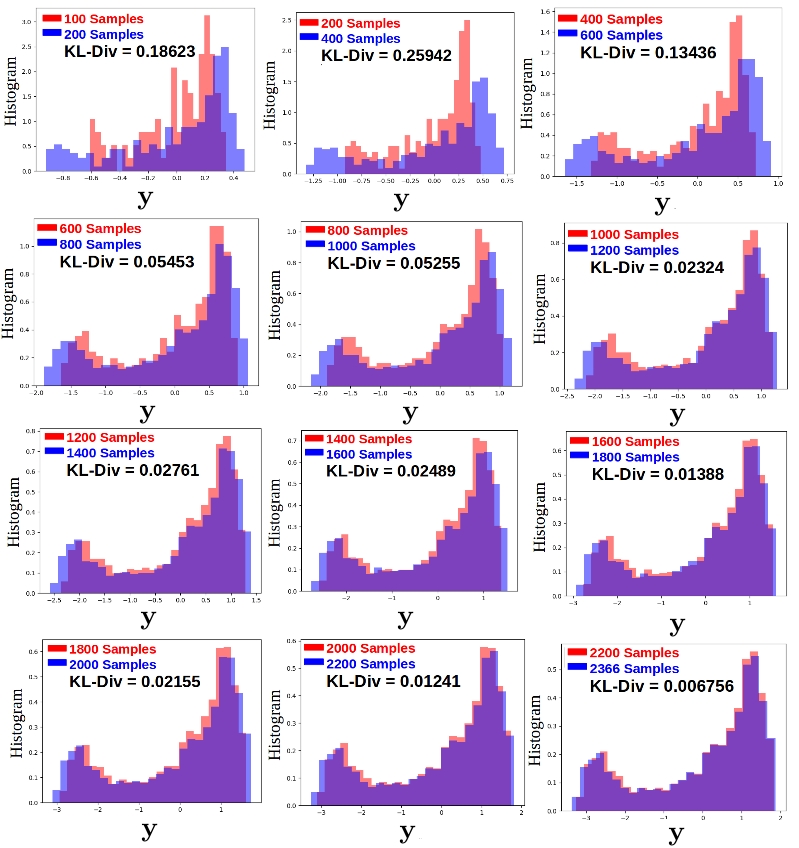}
  \caption{KL divergence evolution between histograms with different sampling size.}\label{KPCA_evo}
 \end{figure}

\begin{figure}[h!]
  \centering
  \includegraphics[width=0.5\textwidth]{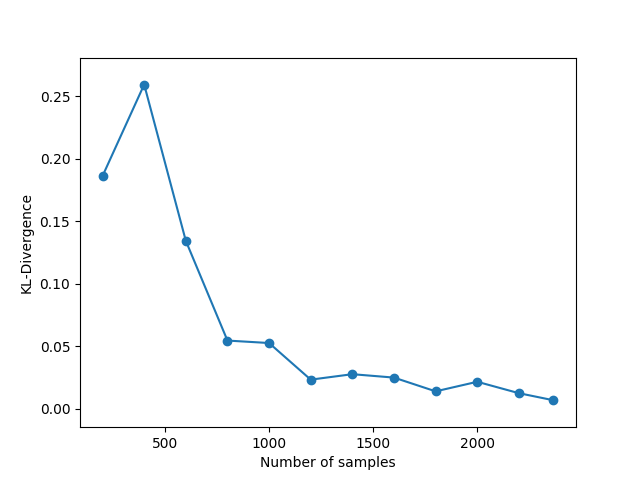}
  \caption{Evolution of KL divergence with respect the number of simulations.}\label{KL_div_evo}
 \end{figure}

 \AGGText{At this point, the number of samples $\ns$  is required to guarantee some statistical accuracy. 
 The KL divergence is used to compare the subsequent distributions of probability of the QoI, obtained with the training sets corresponding to the different values of $\ns$, see  Fig. \ref{ns_approach}.
 That means, to validate the final value of $\ns= 2366$, to be used in the input matrix $\mathbf{X}$ for analysis (instead of the starting value of $\ns = 100$).
 That is, the histogram obtained by the values of $\mathbf{y}$ for a low number of samples $\ns$ is compared by the KL divergence criterion with the histogram for a higher number of samples. This process is repeated by increasing the number of samples until the value of the KL divergence becomes smaller than $10^{-2}$ (considered low enough for our required accuracy). The results obtained in this process are detailed in Fig.\ref{KPCA_evo}, where it is clear how the histograms become more stable increasing the number of samples, and consequently the KL-Div value decreases. For the final number of samples $\ns = 2366$, the KL-Div value is below the prescribed tolerance. Recalling expression \eqref{eq:KLDivDisc0}, the calculated value of $D_{KL}^{0}$ for the final histogram in Fig.\ref{KPCA_evo} is $D_{KL}^{0} = 0.2361$. Thus, the relative value of the difference of the last two distributions is of $2.9\%$.
 
 Additionally, Fig.\ref{KL_div_evo} shows how the KL-Div value becomes stable when the number of samples is rich enough. At this point, increasing the number of samples does not add extra information to the model.}

\begin{figure}[h!]
 \centering
 \subfloat[]{
    \includegraphics[width=0.5\textwidth]{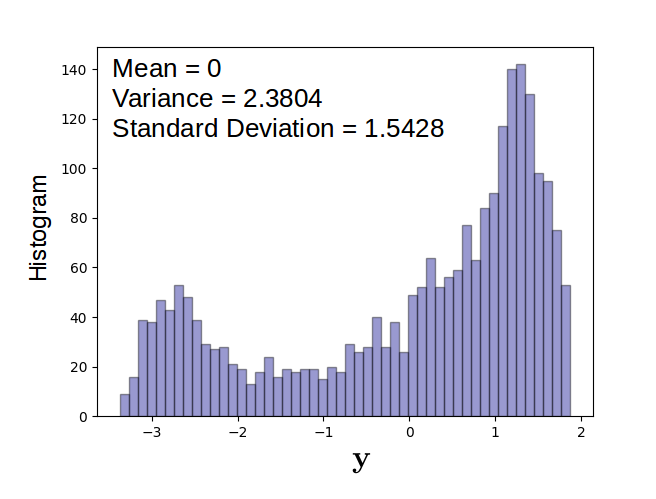}}
 \subfloat[]{
 \includegraphics[width=0.5\textwidth]{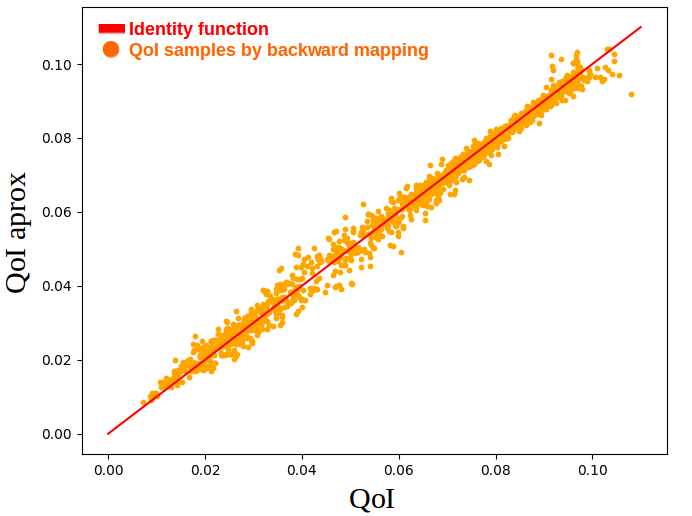}}
  \caption{(a) Reference values of histogram, mean, variance and standard deviation of the first principal component $\mathbf{y}$ of kPCA. This reference values are achieved with 2366 simulations in VPS/Pamcrash. (b) Scatter plot around the identity function (red) of the $QoI$ with respect to the approximated for the first principal component $\mathbf{y}$.}\label{backward}
\end{figure}

 \AGGText{Fig.\ref{backward}a shows the final histogram obtained for the final sampling, $\ns = 2366$. Additionally, the low dispersion of the results by using one reduced dimension is confirmed (the first eigenvalue contains $82.61\%$ of information, stored in $\mathbf{y}=[y^1y^2...y^{\ns}]^T$). Moreover, the consistency of  the \emph{backward mapping} from vector $\mathbf{y}$ to  $\mathbf{x}$ and then calculating the corresponding QoI is also confirmed by the results in Fig.\ref{backward}b.}

\subsection{Link between input parameters and Principal Components}\label{subsec:linking}

\begin{figure}[h!]
  \centering
  \includegraphics[width=0.7\textwidth]{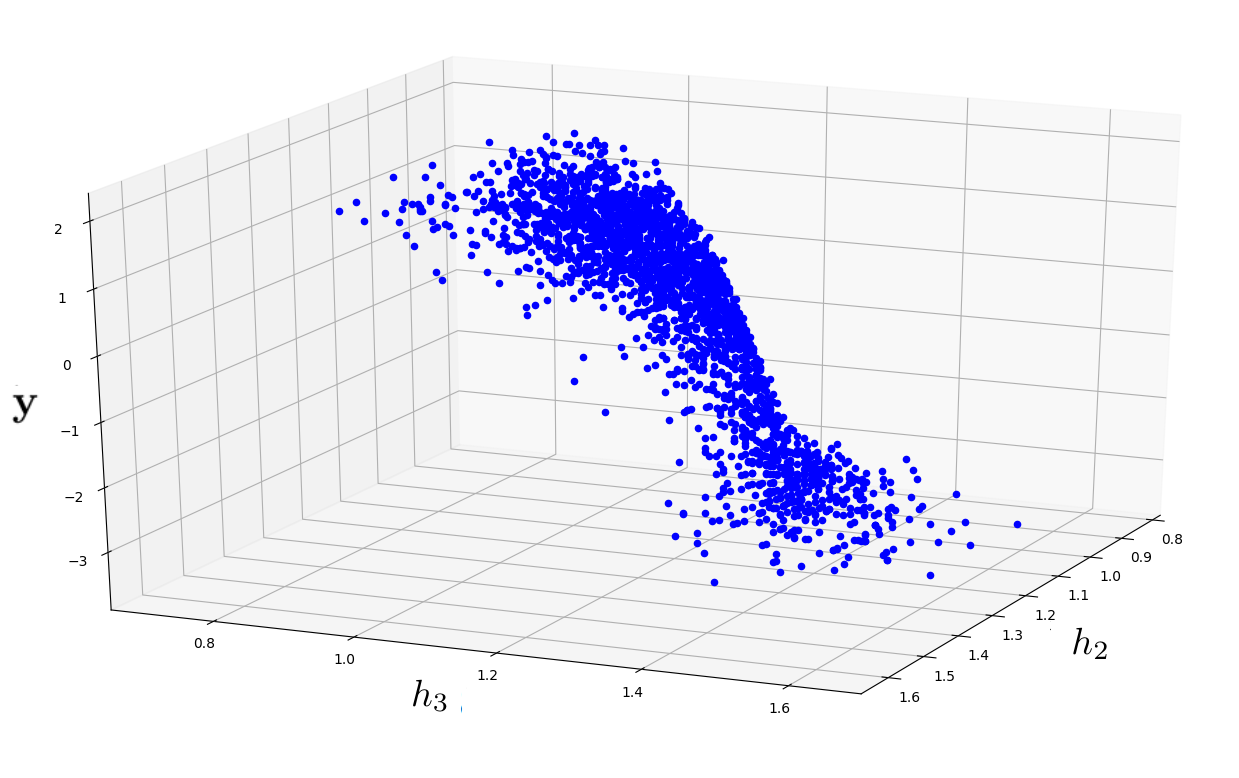}
  \caption{Scatter plot between the reduced space $\mathbf{y}$ and the inputs $h_2,h_3$.}\label{scatter}
 \end{figure}

The first principal component $\mathbf{y}$ is linked to its corresponding values of $\bx=[h_1,h_2,h_3]^T$. Fig. \ref{scatter} shows the scatter plot between the reduced space and the inputs $h_2,h_3$. Moreover, two clusters with different density are observed. The input $h_1$ is discarded by the criterion of Spearman Correlation coefficient (SpC) \cite{hauke2011comparison}. The dependences between the first principal component with respect to the inputs are: 

\begin{itemize}
\item$SpC(\mathbf{y},h_1)=0.035$
\item$SpC(\mathbf{y},h_2)=0.163$
\item$SpC(\mathbf{y},h_3)=-0.968$
\end{itemize}

Clearly, $h_1$ shows a very small correlation, and therefore is discarded for surrogate modeling.

\subsection{Surrogate modeling}\label{subsec:surrogate_modeling}

\begin{figure}[h!]
 \centering
 \subfloat[]{
    \includegraphics[width=0.60\textwidth]{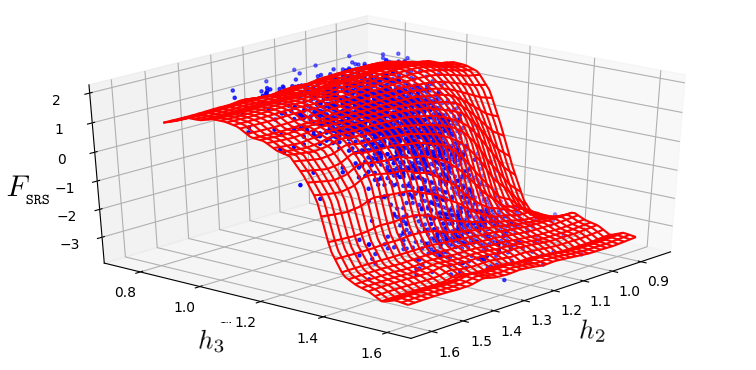}}\\
 \subfloat[]{
 \includegraphics[width=0.62\textwidth]{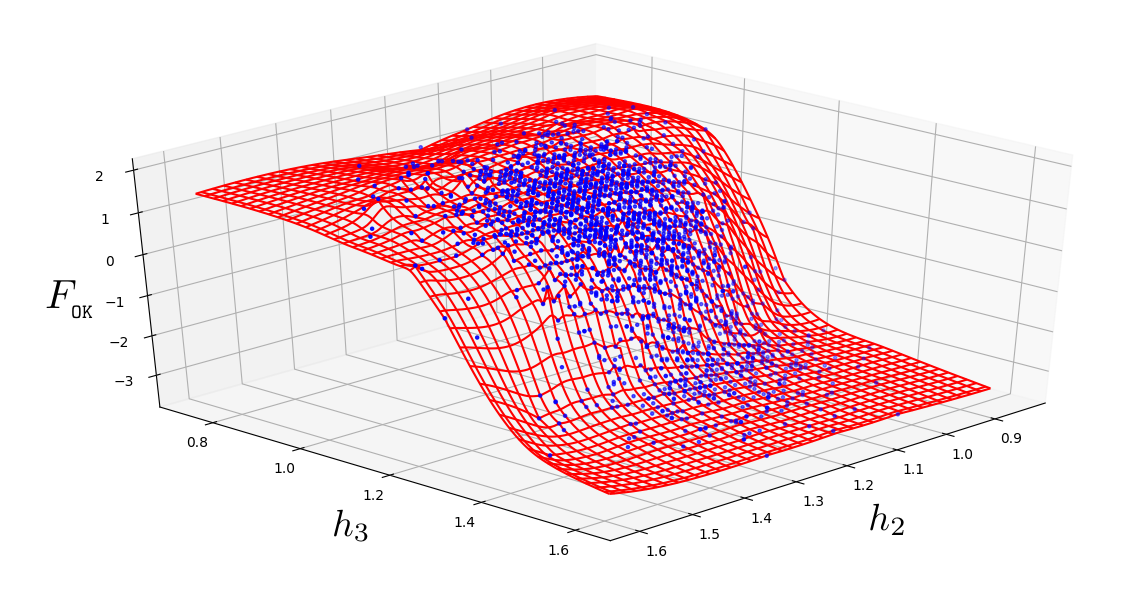}}\\
    \subfloat[]{
    \includegraphics[width=0.6\textwidth]{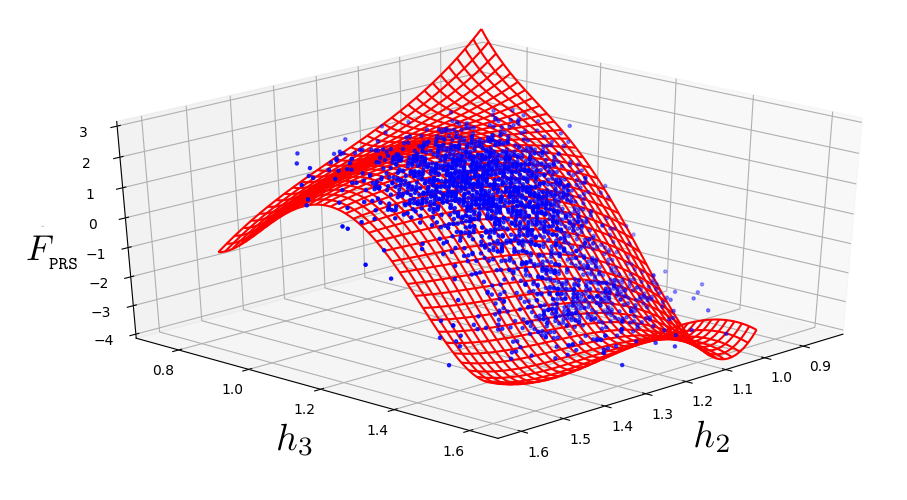}}\\
  \caption{In red, is shown the response surfaces of a) SRS, b) OK and c) PRS. In blue, the scattering samples.}\label{PRS_metamodel}
\end{figure}

In Fig.\ref{PRS_metamodel} it is shown the response surface of SRS, OK and PRS \MRAText{metamodels} between the first principal component of kPCA (\MRAText{storing $82.61\%$ of information}) and the inputs $h_2,h_3$. \MRAText{The metamodels $ F_ {\OK} $ and $ F_ {\SRS} $ show adaptive behaviour with the sample points (in blue). However, the $ F_ {\PRS} $ metamodel exhibits unstable tails in areas where there are few samples from the training set. This problem is called Runge's phenomenon and is common due to lack of sampling in the tails of the distributions.} To evaluate the behaviour and the robustness of the response surfaces, each surface is evaluated increasing new random samples until 50000 points, aiming to compare histograms, means and standard deviations with respect to the reference values plotted in Fig.\ref{backward}a.

\begin{figure}[h!]
 \centering
 \subfloat[]{
    \includegraphics[width=0.6\textwidth]{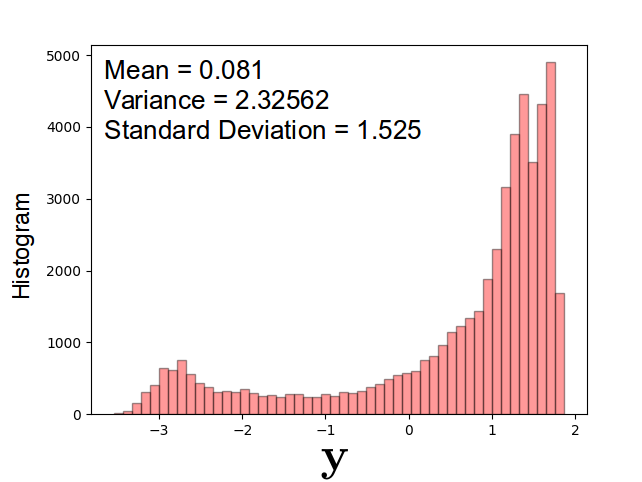}}\\
 \subfloat[]{
 \includegraphics[width=0.6\textwidth]{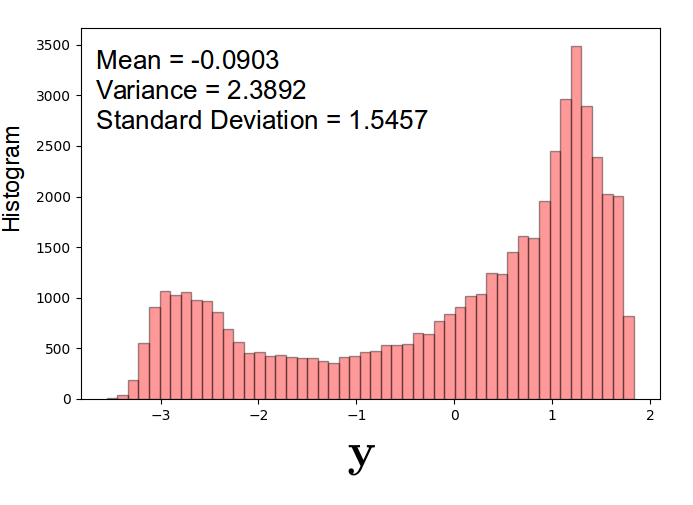}}\\
    \subfloat[]{
    \includegraphics[width=0.6\textwidth]{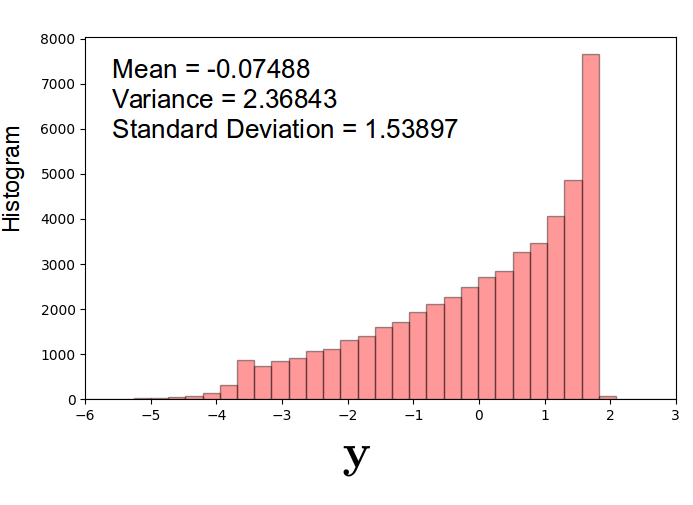}}\\
  \caption{Histogram, mean and standard deviation results of the different surrogate models. a) SRS, b) OK and c) PRS. The results are obtained by evaluating each surrogate model with 50000 new random samples.}\label{metamodel_results}
\end{figure}

In Fig.\ref{metamodel_results} the histogram, mean, variance and standard deviation results of 50000 new random samples are illustrated for the evaluation of each metamodel. \MRAText{The three metamodels give approximations to the mean with an error around 5\% of the Standard Deviation. Since the current reference mean (Figure \ref{backward}a) is zero, both positive and negative values can be expected. The histograms of the $ F_{\SRS} $ and $ F_{\OK} $ metamodels show similar bimodal distribution with respect to the reference histogram illustrated in Fig. \ref{backward}a. However, the two modes of distribution are not captured with the $F_{\PRS}$ metamodel. The Runge's phenomenon of the response surface and a worse adaptation to the sampling points overlooks the two distribution modes.}

\begin{figure}[h!]
 \centering
 \subfloat[]{
    \includegraphics[width=0.6\textwidth]{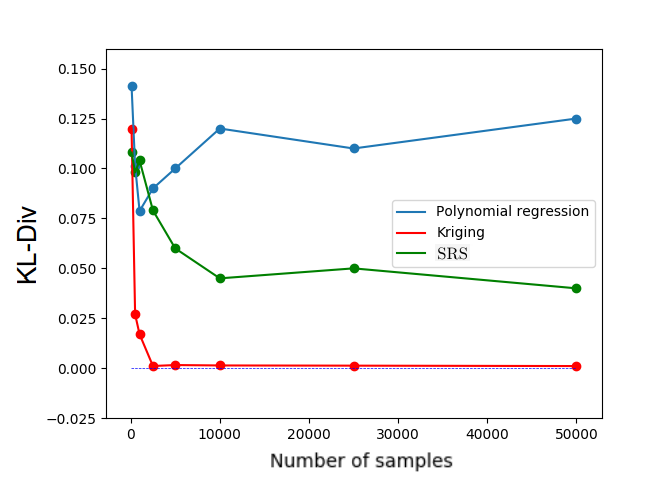}}
 \subfloat[]{
 \includegraphics[width=0.6\textwidth]{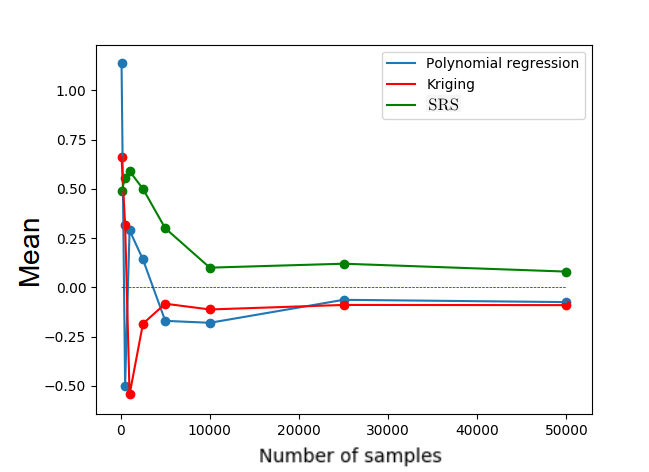}}\\
    \subfloat[]{
    \includegraphics[width=0.6\textwidth]{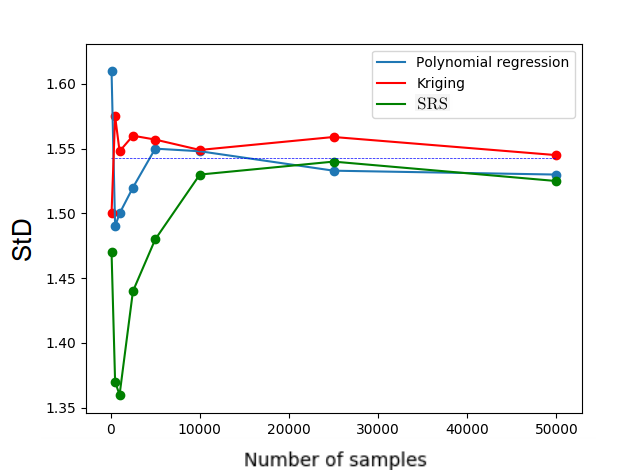}}\\
  \caption{Convergence plots of SRS, OK and PRS evaluating KL divergence, mean and standard deviation with respect to the reference values(KL=0, mean=0, standard deviation=1.5428 plotted with the dashed line).}\label{metamodel_comparative}
\end{figure}

In Fig. \ref{metamodel_comparative} the convergence of the three surrogate models are compared with the reference values while new random points for each metamodel are increased up. The results of this comparative study shows similar results in terms of mean and standard deviation for the three techniques. Meaning that these statistical values are not sensitive for the criterion to select the best surface. However, the results for the KL divergence clearly shows a worse behaviour for the PRS method \MRAText{caused for the tails of the response surface}. In contrast, SRS and OK have similar results for the KL divergence where a very good performance is observed.

\subsection{Uncertainty quantification for the surrogate model}\label{subsec:preimages}

\begin{figure}[h!]
\centering
 \subfloat[]{
    \includegraphics[width=0.6\textwidth]{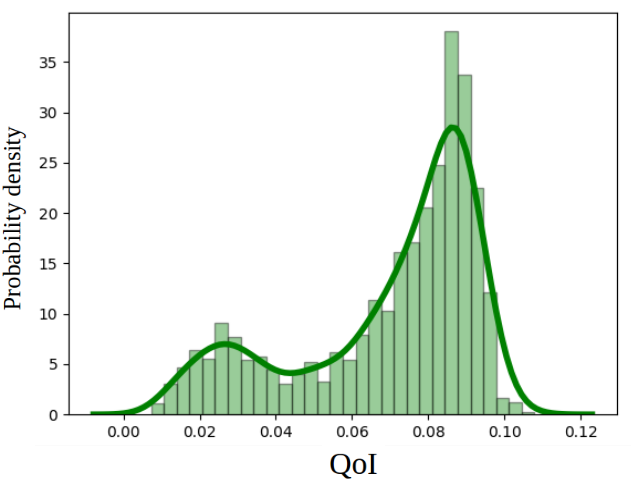}}
 \subfloat[]{
 \includegraphics[width=0.58\textwidth]{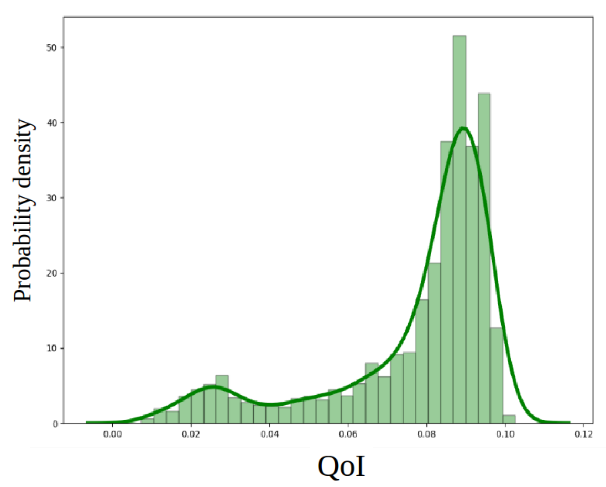}}\\
 \subfloat[]{
	\includegraphics[width=0.6\textwidth]{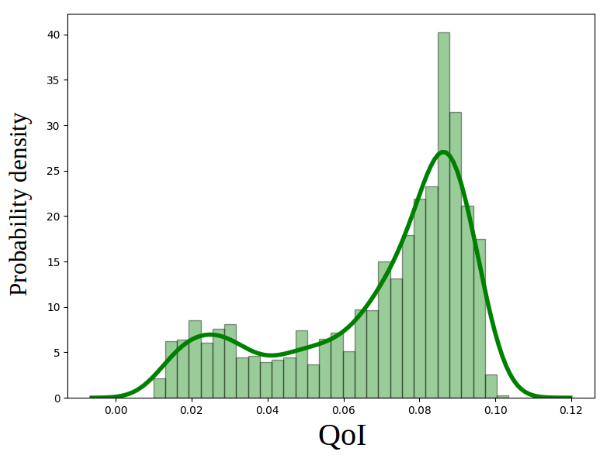}}
 \subfloat[]{
	\includegraphics[width=0.58\textwidth]{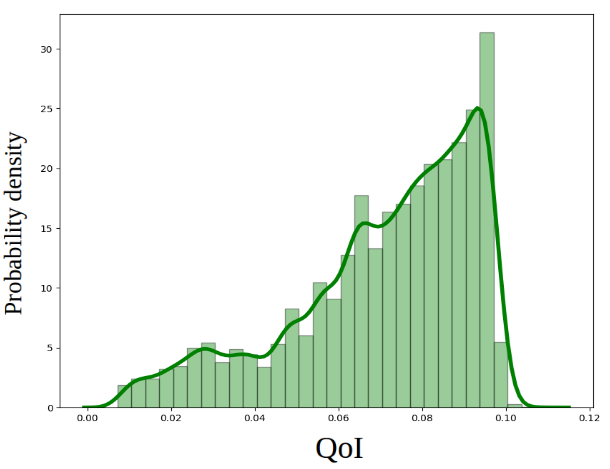}}
\caption{a) shows the histogram of the 2366 reference samples. b), c) and d) shows the histograms of the QoI by evaluating 50000 random samples for SRS, OK and PRS metamodels.}\label{qoi}
\end{figure}

\begin{figure}[h!]	
\centering
\subfloat[]{
\includegraphics[width=0.65\textwidth]{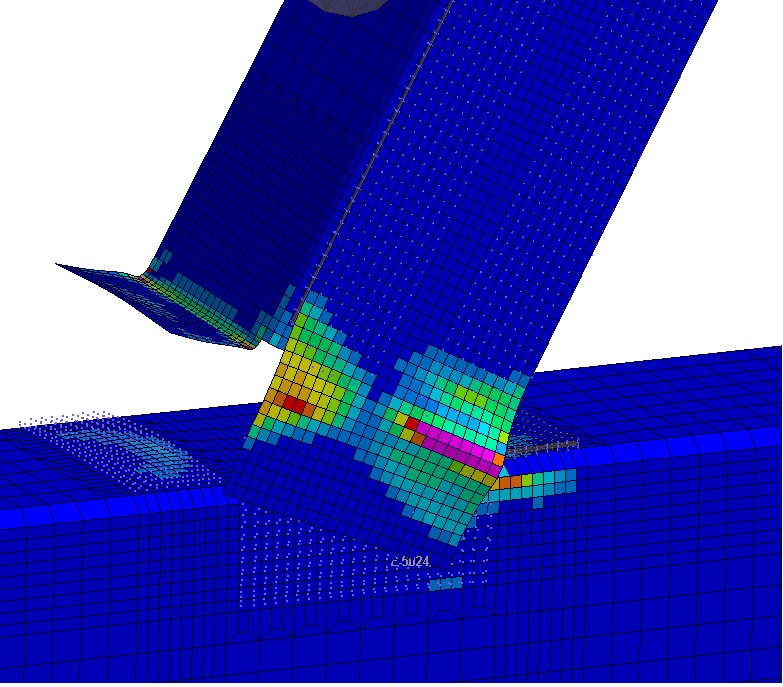}}\\
\subfloat[]{
\includegraphics[width=0.65\textwidth]{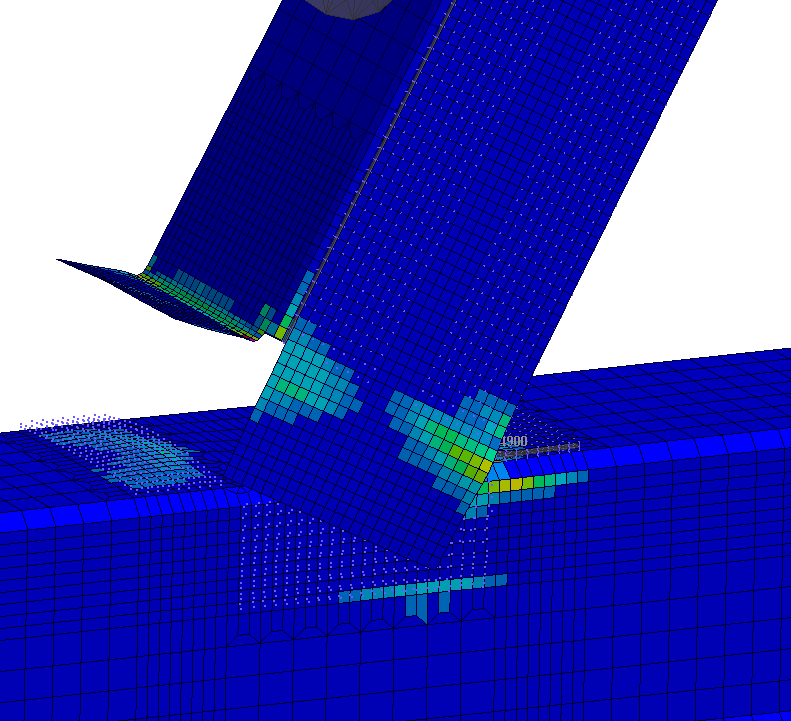}}\\
\caption{a) Snapshot simulation of the  plastic strain in the biggest mode in the QoI histogram. b) Snapshot simulation of the plastic strain in the smallest mode in the QoI histogram.}\label{simulations}
\end{figure}

\MRAText{Once the surrogates $F(\cdot)$ are available, for each input value $\bh$, the corresponding $\mathbf{z^{\star}}$ is straightforwardly computed as $F(\bh)$. Then, the backward mapping produces the corresponding input vector $\mathbf{x}$ of plastic deformation values in the area of interest, bieng $l^{0}(\mathbf{x})$ its the associated QoI. The concatenation of the three operations is computationally negligible with respect to the cost of the training set of full order simulations. At this point, standard Monte Carlo is performed with 50000 new random samples for  $h_1,h_2$ and $h_3$ to evaluate each metamodel. In Fig. \ref{qoi} it is presented the corresponding PDFs of the QoI for the metamodels (SRS, OK and PRS). A bimodal function with approximately $19\%$ of probability for the small mode and $81\%$ for the big mode can be appreciated for SRS and OK. Otherwise, PRS fails to capture such behaviour. The statistical variables of the QoI for each metamodel are presented in Table \ref{tab:table1}. Here the three variables present similar results, which means that any metamodel captures similar information in terms of mean, variance and standard deviation. 

Recalling that kPCA improves the mapping back to the original variable $\mathbf{x}$, a physical interpretation of the bimodal PDF can be performed. The corresponding behaviour of the structure for each mode of the PDF is illustrated in Fig. \ref{simulations}. Clearly, two physical modes are observed. The first snapshot of Fig. \ref{simulations} shows the higher values of plastic strain and a significant back-bend of the plate profile. Otherwise, the second snapshot shows lower values of plastic deformation with a more rigid behaviour of the plate. The first physical case present $81\%$ of probability and the second case $19\%$ of probability occurrence.}


\begin{table}[H]
  \begin{center}
    \caption{Statistical variables for each metamodel.}
    \label{tab:table1}
    \begin{tabular}{l|c|c|r} 
      &\textbf{Mean} & \textbf{Variance}& \textbf{StD}\\
      \hline
      \textbf{Reference values} &0.0695 & 0.1546 & 0.0239 \\
      \hline
      \textbf{SRS} &0.07245 & 0.1452 & 0.0211\\
       \hline
      \textbf{OK} &0.0659 & 0.15 & 0.0225\\
       \hline
      \textbf{PRS} &0.0707 & 0.1479 & 0.0219\\
      
    \end{tabular}
  \end{center}
\end{table}

\section{Discussion and Conclusions}\label{sec:conclusions}
A nonintrusive methodology to perform Uncertainty Quantification for crashworthiness problems is presented. The basic idea is to combine some dimensionality reduction technique (here kPCA) with a surrogate model based on a training set of full-order solutions (ideally not too many, because of their computational cost). The dimensionality reduction eases the task of the surrogate model and enables the analyst to detect clusters and categorize the data.  The surrogate model (or metamodel) substitutes at a negligible computational cost the original full-order model. It therefore permits producing multiple queries to the model, corresponding the different parametric input values demanded by the Monte Carlo strategies.

\AGGText{In the benchmark problem considered, kPCA allows describing the full phenomenon with only one principal component, accounting for more than $82\%$ of the total variance (that is, of the information). This problem is relevant in automotive engineering (and often used as benchmark by SEAT engineers) and, despite the fact that only three input parameters are assumed to have stochastic nature (and 3 dimensions are not awakening the \emph{curse of dimensionality}), the dimensionality reduction is still pertinent to simplify the output of interest to be analyzed. Actually, using kPCA, only one principal component is accounting for more than $82\%$ of the total information. It also detects two clusters corresponding to two deformation modes and two different levels of the QoI. The UQ methodology is also providing the probabilities of occurrence of these two modes, which are $19\%$ and $81\%$. This is reflected in a bi-modal PDF, one mode having a probability four times larger than the other. Moreover, using kPCA as dimensionality reduction strategy the backward mapping from the reduced space is more accurate and allows interpreting the mechanisms associated with these two modes.}



\AGGText{In the presented methodology for a crash problem, the use of linear PCA is also a suitable option. On the one side, if a single scalar QoI is required for decision making, PCA is  simpler than kPCA. On the other side, if it is necessary to find both an accurate QoI, as well as a more detailed approximation of the full original variable $\mathbf{x}$ map, corresponding to a complete deformation field, kPCA improves the mapping back to the original input space, since it accounts on the intrinsic nonlinearities involved in the manifold of training set. Thus, both PCA or kPCA can be used for this methodology, depending on the main objective sought. In this manuscript, for the reasons mentioned above,  kPCA is used and described in more detail. This allows dealing with problems representing more complex phenomena, where data lies in highly nonlinear manifold.}

Three formats of the surrogate models are taken into consideration, Ordinary Kriging (OK), Polynomial Response Surface (PRS) and a Separated Response Surface (SRS) approach, introduced here and based in the PGD methodology. The assessment of the mean and variance of the outcome (the QoI) is properly computed using the three alternative surrogates. However, when it comes to analyze the PDFs (approximated by histograms), the SRS and OK surrogates perform much better than the PRS. The PRS surrogate fails to capture the bi-modal character of the PDF.

Being OK an interpolative methodology (the response surface passes through the data of the training test), it is pretty sensitive to the noise contained in the data. In the current examples, this is not an important issue, because the data is not particularly noisy. However, it may be relevant in other cases. SRS being a least-squares fitting it is not suffering of this drawback.  Moreover, SRS is proposing an explicit parametric solution, therefore it can be used to compute derivatives or to integrate it analytically. This allows also to compute the statistical moments, probability density function and cumulative density function with analytical methods, circumventing the Monte Carlo sampling. Another interesting feature of the SRS is its fair scalability with the number of input parameters (stochastic dimension). 

The combination of the kPCA manifold learning technique with the different surrogates offers an attractive  framework to perform UQ in complex problems. Here, the application to parametric crashworthiness simulations opens new perspectives. The available alternatives for the surrogates (in particular OK and SRS) and the dimensionality reduction techniques at hand, are a powerful toolbox allowing to attack challenging problems in science and engineering. 
\AGGText{The combination of dimensionality reduction and surrogate models produces accurate solutions at an affordable computational cost, accounting also for the uncertainty, that is assessing the credibility of the simulation. Particularly in the context of crashworthiness UQ, the computational cost is a key issue and a driving force for the research developments in the field. Obviously, increasing accuracy requires a higher computational effort. Finding a trade-off between these two factors is a daily concern for research engineers. This paper intends to provide tools to achieve accurate and credible crashworthiness industrial simulations at an acceptable computational effort.}

\section*{Acknowledgments}
This work is partially funded by Generalitat de Catalunya (Grant Number 1278 SGR 2017-2019 and Pla de Doctorats Industrials 2017 DI 058) and Ministerio de Econom\'ia y Empresa and Ministerio de Ciencia, Innovaci\'on y Universidades (Grant Number DPI2017-85139-C2-2-R).

\section*{Compliance with Ethical Standards}
The authors declare that they have no conflict of interest.

\bibliographystyle{plain} 
\bibliography{Paperbib}





\end{document}